\documentclass[11pt,a4paper]{article}
\usepackage{jheppub}
\usepackage{epsfig}
\usepackage{subfig}
\usepackage[countmax]{subfloat}
\usepackage{booktabs}

\newcommand{\beq}{\begin{eqnarray}}
\newcommand{\eeq}{\end{eqnarray}}

\newcommand {\Zb} {\mathbb{Z}} 
\newcommand {\Rb} {\mathbb{R}} 

\newcommand{\centeron}[2]{{\setbox0=\hbox{#1}\setbox1=\hbox{#2}\ifdim
                                        
\wd1>\wd0\kern.5\wd1\kern-.5\wd0\fi
\copy0

\kern-.5\wd0\kern-.5\wd1\copy1\ifdim\wd0>\wd1
                                       \kern.5\wd0\kern-.5\wd1\fi}}
\newcommand{\ltap}{\>\centeron{\raise.35ex\hbox{$<$}}
                               {\lower.65ex\hbox{$\sim$}}\>}
\newcommand{\gtap}{\>\centeron{\raise.35ex\hbox{$>$}}
                               {\lower.65ex\hbox{$\sim$}}\>}

\newcommand\ZZ{\hbox{\zfont Z\kern-.4emZ}}
\font\zfont = cmss10 

\title{Four tops on the real projective plane at LHC}

\author[a,b]{Giacomo Cacciapaglia,}
\author[a]{Roberto Chierici,}
\author[a]{Aldo Deandrea,}
\author[a]{Luca Panizzi,}
\author[a]{St\`{e}phane Perries,}
\author[a]{Silvano Tosi}

\affiliation[a]{Universit\'e de Lyon, France; Universit\'e Lyon 1, CNRS/IN2P3, UMR5822 IPNL, F-69622 Villeurbanne Cedex, France}
\affiliation[b]{King's College London, Department of Physics, Strand, London WC2R 2LS, UK}

\emailAdd{g.cacciapaglia@ipnl.in2p3.fr}
\emailAdd{chierici@ipnl.in2p3.fr}
\emailAdd{deandrea@ipnl.in2p3.fr}
\emailAdd{panizzi@ipnl.in2p3.fr}
\emailAdd{perries@ipnl.in2p3.fr}
\emailAdd{tosi@ipnl.in2p3.fr}

\abstract{We explore the four top signal $t{\bar t}t{\bar t}$ at the 7 TeV Large Hadron Collider as a probe of physics beyond the 
standard model. Enhancement of the corresponding cross-section with respect to the Standard Model value can 
probe the electroweak symmetry breaking sector or test extra dimensional models with heavy Kaluza-Klein gluons and quarks. 
We perform a detailed analysis including background 
and detector simulation in the specific case of a universal extra--dimensional model with two extra dimensions compactified 
using the geometry of the real projective plane. 
For masses around $600$ GeV, a discovery is possible for an effective cross section above $210$ fb ($36$ fb) for 
1/fb (10/fb) of integrated luminosity.
This implies a branching ratio in tops of the $(1,1)$ heavy photon above $13\%$ ($5\%$). 
Furthermore, the 4-top signal from the $(2,0)$ and $(0,2)$ tiers can be discovered with an integrated luminosity of $3.5$/fb.
The results of our simulation can be easily adapted to other models since the 
background processes are identical. 
Concerning the signal, typical production mechanisms for the $t{\bar t}t{\bar t}$ signal
are similar even if cross-section values may vary considerably depending on the model and the spectrum of the new particles.}

\begin{document}
\begin{flushright}{\tt LYCEN 2011-04}\\
\tt KCL-PH-TH/2011-14
\end{flushright}
\maketitle
\section{Introduction}
\label{sec:intro}
\setcounter{equation}{0}
\setcounter{footnote}{0}

At present the Large Hadron Collider (LHC) is exploring new energy domains and, with increasing integrated luminosity, 
it will soon be capable of discovery or of giving stronger bounds on the type and nature of the new particles expected in 
different extensions of the Standard Model (SM). In the following we propose to explore the four top final state 
$t{\bar t}t{\bar t}$ as a probe of physics beyond the standard model. In the SM, this channel is rather uncommon 
(the cross-section is below 1 fb at the LHC for a 7 TeV centre of mass energy): an enhancement of 
the corresponding cross-section is in principle easy to detect if background processes can be reduced. 
The four top signal has been proposed in the past as a probe of the nature of the electroweak symmetry 
breaking sector at hadron colliders~\cite{Cheung:1995eq,Spira:1997ce}, 
because the coupling of the top quark to the Goldstone bosons and the Higgs particle is of order one. 
Moreover this signal is enhanced in extended or modified electroweak symmetry 
breaking sectors beyond the standard model, such as top condensation models~\cite{Hill:2002ap}, 
models of composite tops~\cite{Pomarol:2008bh}, \cite{Kumar:2009vs},  composite Grand Unified 
Models \cite{Frigerio:2011zg},
and supersymmetric models with light stops and gluinos~\cite{Kane:2011zd}.
In the case of top condensation models, for instance, the top and anti-top quarks are produced by gluon splitting and the 
tops can be strongly scattered via the new interactions of the electroweak symmetry breaking sector. 
Another possible source of enhancement of the $t{\bar t}t{\bar t}$ cross-section appears in extra dimensional models. 
In this framework the four top signal was described as an important probe for low scale warped extra 
dimensions~\cite{Jung:2010ms}. 
The main effect for the enhancement comes, in this case, from a heavy gluon which can be pair produced and give 
the four top quarks or via associate production of the heavy gluon with $t{\bar t}$.
In all these cases the detailed study of the signal and of the background, together with detector simulation, plays a 
major role in understanding the real value of the $t{\bar t}t{\bar t}$ signal in the LHC environment. In the following 
we shall consider a specific extra dimensional model with natural Dark Matter candidate~\cite{Cacciapaglia:2009pa} 
which features a large enhancement of the signal. 
The production of the two top pairs takes place via the resonant decays of vector resonances with masses of a
few hundred GeV. The conclusions of this detailed study can be considered valid in a more general 
context of vector resonances decaying in top pair as long as the signal can be well separated from the backgrounds, 
while a more specific and model dependent study may be required to distinguish this scenario from other new 
physics sources of 4 top events.

The model we will focus on is a 
particular incarnation of Universal Extra Dimensions (UED) on the flat Real Projective Plane (RPP) which is motivated by 
the presence of a Dark Matter candidate as a consequence of the compactification.
It is well known that most of the matter in the universe is dark and of non-baryonic origin. 
Various potential candidates have been 
suggested, and among the most popular in particle physics models there are non-relativistic (massive) and weakly interacting particle (WIMPs).
In this class we find the lightest 
supersymmetric particle (LSP) obtained by imposing R-parity in supersymmetric 
models~\cite{LSP}; the lightest T-odd particle (LTP) obtained by imposing T-parity on little Higgs models~\cite{LTP}; 
the lightest Kaluza-Klein particle (LKP) obtained by imposing a KK parity in UED models~\cite{UED5,LKP}.
The latter case is similar to the one under consideration: the main difference is that in KK parity models, 
it is necessary to impose special conditions on the fixed points of the orbifold for the symmetry to survive; 
on the other hand, on the RPP, a KK parity is a residual of the 6 dimensional Poincar\'e symmetry group 
that is broken by the orbifolding of the two extra space co-ordinates.
In this sense, the parity is a legitimate daughter of the extended Lorentz symmetry, 
therefore we dub the Dark Matter candidate the lightest Lorentz particle (LLP).
A realisation of UED model on the RPP is discussed in \cite{Cacciapaglia:2009pa}, where we refer the reader for 
more details.  The RPP geometry was previously used in \cite{Hebecker:2003we} for building a
6-dimensional grand unified theory.
The main requirement is the absence of fixed points of the orbifold, together with the existence of chiral 
fermions for the zero modes corresponding to the SM particles. These requirements turn out to be quite restrictive, ruling out a model in 5 flat dimensions. 
In 6 dimensions a unique orbifold in flat space satisfies these requirements, the 
Real Projective Plane. 
While the particle content and tree-level interactions are quite similar to other UED scenarios in 6D~\cite{raviolo}, loop corrections, 
which are crucial to study the phenomenology of the model as they are responsible for the masses and decays of the KK resonances, are very different.
In particular, the mass splitting between states in the same KK tier are much smaller.
This has dramatic consequences on the phenomenology of the models: for instance, in decays within the tier, 
the radiated SM particles will have very limited kinetic energy available, therefore they are very difficult to trigger on.
This fact will significantly affect, among others, the 4 top signature we are interested in.

The paper is organised as follows: in section \ref{sec:UEDRPP} we briefly describe the motivations and the structure of the extra--dimensional model.
In section \ref{sec:pheno} we discuss the phenomenology of the model and the different decay modes giving rise to the 4 top final state.
In section \ref{sec:simul} we describe the details of the simulation of the four top signal $t{\bar t}t{\bar t}$ at LHC with 7 TeV centre of mass energy and we discuss the analysis of the signal and backgrounds
Finally in section \ref{sec:conclusion} we give the main results of this study and our conclusions.

\section{Universal Extra Dimensions on the Real Projective Plane}
\label{sec:UEDRPP}
\setcounter{equation}{0}
\setcounter{footnote}{0}

Orbifolds are quotient spaces of a manifold modulo a finite group. 
Here we are interested in flat space: in this case, we can think of the extra dimensions as an infinite Euclidean space $\Rb^d$, where $d$ is the number of extra dimensions, and the orbifolding procedure as a way of 
covering the whole space repeating a patch (the fundamental domain of the orbifold) by use of the isometries of the 
infinite space itself.
In this sense, there are only 2 one-dimensional orbifolds: the circle $S^1 = \Rb/\Zb$ and the interval $S^1/\Zb_2$.
The circle, defined in terms of periodic repetition of an interval, has neither 
fixed points nor chiral fermions; the interval is the only orbifold with chiral massless fermions, however it possesses two fixed 
points (the boundaries of the interval). The presence of fixed points is not a problem in principle. 
Phenomenologically, divergences that arise in loop calculations will require counter-terms to be localised on the fixed points, 
therefore limiting the predictive power of the theory without invalidating the effective theory approach to extra dimensions.
The main issue related to fixed points is that they break the extended Poincar\'e invariance down to the 4-dimensional one, 
therefore any extra symmetry from the extra dimensional space is broken unless {\it ad-hoc} conditions are imposed on the localised Lagrangians.
This is the case of the KK parity. Requiring the absence of fixed 
points, therefore, has a notable consequence: the DM candidate would be a direct consequence of the 
geometry of the orbifold, without a symmetry to be imposed by hand.
Thus, if we extend our four dimensional world to include one extra dimension there are no compactifications without 
fixed points which allow to obtain chiral fermions in 4D. 
The next step is to consider two extra dimensions. 
In the plane the possible isometries are translations ($t$), mirror reflections, $2\pi/n$ rotations ($r$) with $n=2,3,4,6$ 
and glide-reflections ($g$) which are translations with a simultaneous mirror reflection. 
Using these symmetries, in 2 (extra-)dimensions one can define  17 crystallographic symmetry groups, also called the 
wallpaper groups, each giving rise to a different orbifold.
Only 3 of the resulting orbifolds are free of boundaries or fixed points/lines. They are the torus, the Klein bottle and 
the Real Projective Plane. 
Among them, the real projective plane alone allows chiral zero modes for fermions.
The real projective plane is non-orientable and has no boundaries. 
In flat space, it can be thought of as a rectangular patch of a torus with the opposite sides identified as twisted, like a double
M{\"o}bius strip.
The corners of the fundamental rectangle are special points of the space: they are not fixed, 
because the glide identifies the facing pairs, however they are conical singularities with deficit angle $\pi$.
Therefore, localised Lagrangians are allowed on the two singular points, however the identification operated by the glide 
allows to define a preserved KK parity without any further restriction on the localised interactions~\cite{Cacciapaglia:2009pa}. 
The Real Projective Plane can also be obtained from a sphere with identified antipodal points~\cite{Dohi:2010vc}: 
in this case, there are no singular points and the curvature is uniformly spread in the space.
The spectra in the two realisations are very different and, in particular, due to the curvature, fermions are massive on the sphere, 
unless some special mechanisms are in action.

In the following, we will consider 
the former option, as discussed in \cite{Cacciapaglia:2009pa}: we focus on a simple extension of the Standard Model 
on the real projective plane, however similar conclusions will apply to most models based on this background. 
The lowest order Lagrangian is the SM one, extended to 6 dimensions, 
while dimension 6 operators localised on the two singular points are required to absorb one-loop divergences.
To each SM fields it corresponds a tower of massive resonances organised in tiers, 
labelled by two integers $(l, k)$ which correspond to the discretised momenta along the extra directions. 
The field content of 
each tier crucially depends on the parities of the fields under the orbifold symmetries. At leading order, all the states in each tier 
are degenerate with mass determined by the two integers
\beq
m^2_{l,k} = \frac{l^2}{R_5^2} + \frac{k^2}{R_6^2},
\eeq
where $R_{5,6}$ are the radii of the two extra dimensions.
Mass differences within the modes in each tier can be generated by three mechanisms: the Higgs vacuum expectation value (VEV), loop corrections from bulk interactions and higher order operators localised on the singular points.
Due to the flatness of the metric along the extra coordinates, the Higgs VEV is constant and, due to the orthogonality of the 
wave functions of KK modes from different levels, the Higgs mechanism will not mix them.
Therefore, the KK expansion remains valid and the masses will be shifted, independently on the spin of the field, 
by the appropriate SM mass $m_0$; in absence of mixing with other fields, the corrected masses are given by the formula
\beq
m^2_{l,k} = \frac{l^2}{R_5^2} + \frac{k^2}{R_6^2} + m_0^2.
\eeq
On the other hand, the loop corrections generate level-mixings: when diagonalising the mass matrix, however, the off-diagonal 
terms will only contribute at second order, therefore we can consistently limit ourselves to diagonal contributions~\footnote{This is 
not true for tiers that are degenerate at tree level: for instance the tiers $(2,0)$ and $(0,2)$ in the case of degenerate radii.}.
Note also that the loop induced terms will respect the full global symmetries of the bulk space, and therefore, as an example, 
no splitting and/or mixing between the $(1,0)$ and $(0,1)$ levels will be induced, neither decays of the $(1,1)$ modes 
into SM particles. Localised terms are required as counter-terms for the loop divergences: however, they respect less 
symmetries than the bulk loops. 
The only unbroken symmetry will be the KK parity $p_{KK}$.
In the following we will assume that localised terms are small as they correspond to higher dimensional operators in the 6D theory.
Therefore, we will neglect their effect if tree-level or loop contributions are present.
The explicit calculation of the loops shows that in this scenario, contrary to other UED models in the literature, mass differences 
within each tier will be small.

The KK mass, and the Higgs mass, are the main free parameters of the model: calculating the relic Dark Matter abundance 
in this model, one can pin down the cosmologically interesting range for the KK mass.
However, this is nothing but an estimate, because the result is very sensitive to the model of Cosmology and values of 
the cosmological parameters.
The result for the relic abundance as a function of the KK mass was given in~\cite{Cacciapaglia:2009pa} and updated in~\cite{jeremie,relicDM}.
We can consider two limiting cases: 
one in which the two radii are the same so that two degenerate tiers $(1,0)$ and $(0,1)$ contribute to the Dark Matter abundance.
In this case, the preferred mass range is
\beq
200\;\mbox{GeV} < m_{KK} < 300\;\mbox{GeV}\,;
\eeq
and a limit $m_{KK}<400$ GeV from the over-closure of the Universe.
The two degenerate tiers may be split in the case of asymmetric radii: when one radius is smaller than the other by more than few 
percent (in particular if the difference in mass is larger that the freeze out temperature, which is typically of order 4\% of the KK 
mass) the heavier tier does not contribute significantly to the relic abundance and the range is
\beq
300\;\mbox{GeV} < m_{KK} < 400\;\mbox{GeV}\,.
\eeq
We will consider in the following the case $m_{KK} = 400$ GeV in detail.
The  tiers $(1,0)$ and $(0,1)$ will be considered therefore of a mass of 400 GeV (apart from small splitting corrections) 
and the lightest of them will be associated to the dark matter candidate. Due to the the small mass splittings among the 
particles inside the level $(1,0)$ and $(0,1)$, the typical phenomenology at LHC is the production of coloured particle 
via tree level bulk interactions. Due to the conservation of KK-parity, these states will be pair-produced and will decay into the LLP, 
which is a scalar photon, after radiating soft standard model particles. Indeed due to the small mass splittings, the energy 
available for the standard model particle is small and they are soft. 
Therefore, a very detailed study is necessary to ascertain if missing energy signatures can be detected at the LHC.

Here, we will focus on the phenomenology of heavier states, especially the next two in mass, 
$(1,1)$ and $(2,0)$--$(0,2)$, which are even under the KK parity and can therefore decay in a pair of SM particles. 
Despite the lower cross sections due to the heavier mass scale of these tiers, interesting signals can be generated, 
as will be described in the following sections.

\section{Phenomenology of the (1,1) tier}
\label{sec:pheno}
\setcounter{equation}{0}
\setcounter{footnote}{0}

The tier $(1,1)$ is peculiar to models with more than one extra dimensions and it has been pointed as a signature of models beyond 5D~\cite{Burdman:2006gy}.
Its tree-level mass is given by $M^{(1,1)} = \sqrt{1/R_5^2 + 1/R_6^2} \sim \sqrt{2}\, M_{KK}$, where the latter is true if the two radii are nearly equal.
In this case, this level is the next-to-lightest KK tier after the $(1,0)$ and $(0,1)$.
Under the KK parity, this level is even and, due to its mass, it cannot decay into a $(1,0)$-$(0,1)$ pair via bulk interactions: the only possibility is a direct coupling to a pair of SM states.
Such a coupling, however, cannot be generated by loops of bulk interactions due to the conservation of KK momentum in the vertexes.
The decay, therefore, can only occur via localised counter-terms that break all parities but the KK parity.
In the context of a valid effective theory, we assumed that such couplings are small, corresponding to higher order operators, and a simple dimensional analysis shows that they are typically much smaller than loop induced couplings.
This means that any state in this tier will preferably chain-decay to the lightest one in the tier.
Such decays are in fact ruled by the phase space, which is in turn determined by loop and Higgs VEV corrections.
The fate of the lightest state is somewhat arbitrary: according to the magnitude of the higher order operators, it may be long lived and escape the detector, leave a displaced vertex or decay promptly into SM particles.
However, based on dimensional analysis,  an estimate of the size of the coupling as suppressed by the low cut-off of the theory would give rise to prompt decays, so we will focus on this possibility in the following.

\subsection{Spectrum and interactions}
\label{subsec:spectrum}

The tier $(1,1)$ has a rich spectrum compared to other tiers: it contains gauge vectors ($A_{\mu}^{(1,1)}$) and gauge scalars ($A_{5,6}^{(1,1)}$), Dirac 
fermions ($f^{(1,1)}$) and Higgs ($H^{(1,1)}$). For each SM gauge boson, for instance the $W^-_\mu$, the spectrum contains a 
vector resonance $W^{-,(1,1)}_\mu$ and a physical gauge scalar $W_\phi^{-,(1,1)}$, which is the combination of the extra polarisations 
that is not eaten by the massive vector resonance.
The masses are split by Higgs VEV and loop corrections: in particular, loop corrections are divergent for the vector resonances 
and finite (and therefore very small) for the gauge scalars.
This is due to the fact that the wave functions of the gauge scalars vanish on the singular points where the counter-terms are 
located. The Higgs resonance $H^{(1,1)}$, on the other hand, will receive quadratically divergent corrections, therefore it is 
difficult to predict its mass. However, as a general rule, it will tend to be heavier that other states in the tier, therefore its presence 
will be irrelevant for the purpose of this work.
In the fermion sector, the spectrum contains 2 massive Dirac fermions for each chiral SM fermion: this means that for each SM 
fermion, for instance the electron $e$, there will be 4 massive states: $e_s^{a,(1,1)}$, $e_s^{b,(1,1)}$, $e_d^{a,(1,1)}$ and 
$e_d^{b,(1,1)}$.
The indices $s$ and $d$ refer to singlet (right-handed SM electron) and doublet (left-handed SM electron), while $a$ and $b$ 
label the two degenerate states deriving from the KK expansion.
Singlet and doublet fermions will be distinguished by the interactions (for instance, they will receive different loop corrections), while 
the difference between fermions $a$ and $b$ is more subtle.
One way to distinguish them is to define by $a$ the fermion that receives divergent loop corrections, while $b$ receives finite 
corrections (some more details about this can be found in~\cite{Cacciapaglia:2011hx}).

Radiative corrections are crucial to understand the phenomenology of every tier, because, together with the Higgs VEV, they 
determine the mass difference between states in the tier.
In the case of the $(1,1)$ tier, this is directly related to the decay channels of each state.
Details on techniques to compute loop diagrams in 6D can be found on \cite{Cacciapaglia:2009pa}: the main novelty in the case 
under consideration is the fact that we are dealing with states which have non-zero KK numbers
for both extra directions and therefore the wave-function contains both sine and cosine functions.
To obtain the spectrum of the tier, one can either calculate directly the loops (the result is reported in Appendix~\ref{app:calcul}), 
or use the counter-term structure as in~\cite{Cacciapaglia:2011hx}. 
The spectrum is clearly divided into a set of states that receive largish UV-sensitive 
corrections (vector resonances and fermions of type $a$), and a set that receive small finite corrections (gauge scalars and 
fermions of type $b$).

As mentioned before, the decays of the states in tier $(1,1)$ are mediated by bulk interactions that contain two (different) states in 
the level $(1,1)$ and one (or more) SM particles.
Here the separation into two sets appears again, as there are no vertexes connecting the two sets.
In fact, fermions of type $a$ can only interact with a SM fermion via a vector resonance of level $(1,1)$, while fermions of type $b
$ only interact with a gauge scalar.
This fact can be understood looking at the wave-functions of the heavy states \cite{Cacciapaglia:2009pa}. 
The overlap of the wave-functions in a vertex must contain an even number of $\cos$ and $\sin$ functions in order 
for the integral on the extra-coordinates to be non-vanishing.
The wave functions of the gauge boson components are proportional to:
\begin{equation*}
 A_{\mu}^{(1,1)} \sim \cos l x_5 \cos k x_6\,, \quad A_5^{(1,1)} \sim \sin l x_5 \cos k x_6\,, \quad A_6^{(1,1)} \sim \cos l x_5 \sin k x_6\,.
\end{equation*}
Fermions in 6D are eight-component spinors whose wave-functions are different whether the zero mode is left- or right-handed. If the zero mode is left-handed, for instance, the 4D components are:
\begin{eqnarray*}
 P_Lf_d^{a,(1,1)} \sim \cos l x_5 \cos k x_6\,, && P_Rf_d^{a,(1,1)} \sim a_1 \cos l x_5 \sin k x_6 + a_2 \sin l x_5 \cos k x_6\,, \\
 P_Lf_d^{b,(1,1)} \sim \sin l x_5 \sin k x_6\,, && P_Rf_d^{b,(1,1)} \sim b_1 \cos l x_5 \sin k x_6 + b_2 \sin l x_5 \cos k x_6\,.
\end{eqnarray*}
A vertex of the type $A_\mu^{(1,1)}\bar{f}^{(1,1)}f^{(0,0)}$ requires that the two fermions have the same chirality:
 $f^a$ has such couplings since it provides an even number of $\cos$ and $\sin$ functions (the wave function of the SM particle is constant). 
On the other hand, a vertex of the type $A_\phi^{(1,1)} \bar{f}^{(1,1)}f^{(0,0)}$ flips fermion chirality:
it is possible to prove that the combination of $A_5$ and
$A_6$ which generates the physical gauge scalar $A_\phi$ couples only with fermions $f^b$.
If the zero mode is a right-handed fermions, the 4D components of the $(1,1)$ modes have different combinations of $\sin$ and $\cos$, 
but the analysis is completely analogous and leads to the same conclusions.
In summary, the following interactions are possible:
\begin{eqnarray*}
 A_{\mu}^{(1,1)} \bar{f}_{s,d}^{a,(1,1)} f^{(0,0)} + h.c.\,, \qquad
 A_\phi^{(1,1)} \bar{f}_{s,d}^{b,(1,1)} f^{(0,0)} + h.c.\,,
\end{eqnarray*}
while the following ones vanish:
\begin{eqnarray*}
 A_{\mu}^{(1,1)} \bar{f}_{s,d}^{b,(1,1)} f^{(0,0)}+ h.c.\,, \qquad A_\phi^{(1,1)} \bar{f}_{s,d}^{a,(1,1)} f^{(0,0)}+h.c.\,.
\end{eqnarray*}

We have shown, therefore, that the tier $(1,1)$  contains two distinct sectors which do not interact and can be treated separately: 
$\{f^{a,(1,1)},A_{\mu}^{(1,1)}\}$, that receives divergent corrections to the masses from loop diagrams, and $\{f^{b,(1,1)},A_\phi^{(1,1)}\}$, that receives finite loop corrections.
In both cases, each state will decay to the lightest one in its set via bulk interactions.
In the set $\{f^{a,(1,1)},A_{\mu}^{(1,1)}\}$, such decays are prompt because there is enough phase space available for the decay; 
moreover, it is possible to write down localised operators that will induce the decay of the lightest state, a neutral vector 
resonance, into a pair of SM particles.
Due to the fact that the vector boson wave function does not vanish on the fixed points, all channels are possible: a pair of fermion anti-fermion, $W^+ W^-$ and a pair of Higgs bosons.
From the operator structure alone it is not possible to determine which final state dominates.
A reasonable assumption is that the decay will preferably involve heavy SM particles because they may be more sensitive to the underlying fundamental theory, therefore in the following we will focus on decays into a pair of top anti-top, leaving the value of the branching ratio as a free parameter of the model.
In the case of the set $\{f^{b,(1,1)},A_\phi^{(1,1)}\}$ the mass difference between 
different states will be very small, typically below $1$ GeV, with the sole exception of the $Z^{(1,1)}_\phi$, $W^{\pm,(1,1)}_\phi$ and top partners 
which receive large corrections from the Higgs VEV.
The decay within the level, therefore, will produce extremely soft SM particles that will be virtually impossible to detect.
The lightest state in the set will decay into standard model states via localised operators: however, due to the fact that the wave 
function vanishes on the singular points, the decay will be mediated by operators of dimension 6 or higher, and the width will be 
very suppressed. A detailed discussion is given in Appendix~\ref{app:locops}, where we show that the scalar $A_\phi^{(1,1)}$ decays typically 
in a pair of Higgs bosons while the decay to top pairs is suppressed.
In the following we will focus on the phenomenology of the set containing vector resonances and fermions of type $a$.
For simplicity of notation, from now on we will drop the index $a$ on the fermion.
Masses, widths and branching ratios for the benchmark point with $M_{KK} = 400$ GeV are listed in 
Table~\ref{tab:11spectrumdecaychannels}.

\begin{table}[t!]
\begin{center}
{\footnotesize
\begin{tabular}{ccccr}
\toprule
Particle & $\begin{array}{c}\rm{Mass}\\ \rm{(GeV)}\end{array}$ & $\begin{array}{c}\rm{Width}\\ \rm{(MeV)}\end{array}$ & Decay Channels & $\begin{array}{c}\rm{Branching}\\ \rm{Ratios}\end{array}$ \\
\midrule
\midrule
\multicolumn{4}{c}{Gauge Bosons} \\
\midrule
\midrule
$A_\mu^{(1,1)}$  & 566.39 & $\mathcal{O} (10^{-3})$
& $t\bar t$ & $b$ \\ 
                &         &                     & others ($h h$?)  &  $1-b$ \\
\midrule
$g_\mu^{(1,1)}$  & 661.86 & $4.15\times10^{3}$  & $d~\bar d_s^{(1,1)},\bar d~d_s^{(1,1)}$ & $5.73\%$ \\ 
                 &        &                     & $b~\bar b_s^{(1,1)}, \bar b~b_s^{(1,1)}$                                               & $5.72\%$ \\ 
                 &        &                     & $u~\bar u_s^{(1,1)}, \bar u~u_s^{(1,1)}$ & $5.55\%$ \\ 
                 &        &                     & $d~\bar d_d^{(1,1)}, \bar d~d_d^{(1,1)}$ & $4.46\%$ \\ 
                 &        &                     & $b~\bar b_d^{(1,1)},\bar b~b_d^{(1,1)}$                                               & $3.87\%$ \\ 
                 &        &                     & $u~\bar u_d^{(1,1)},\bar u~u_d^{(1,1)}$ & $4.46\%$ \\ 
\midrule
$Z_\mu^{(1,1)}$  & 589.35 & $36.9$ & $\nu_{l}\bar\nu_{l}^{(1,1)}, \bar\nu_{l}\nu_{l}^{(1,1)}$ & $9.38\%$ \\ 
                 &        &                     & $l^+ l_L^{-,(1,1)}, l^- l_L^{+,(1,1)}$   & $6.98\%$ \\ 
                 &        &                     & $l^+ l_R^{-,(1,1)}, l^- l_R^{+,(1,1)}$   & $0.33\%$ \\ 
\midrule
$W_\mu^{(1,1)}$  & 588.95 & $34.8$ & $l^-\bar\nu_l^{(1,1)},\bar\nu_l l^{-,(1,1)}$ & $16.6\%$ \\ 
\midrule
\midrule
\multicolumn{4}{c}{Quarks} \\
\midrule
\midrule
$u_{s}^{(1,1)}$          & 601.96 & $13.8$     & $u A_\mu^{(1,1)}$   & $99.8\%$ \\ 
                               &        &                         & $u Z_\mu^{(1,1)}$        & $0.2\%$  \\
\midrule
$t_{s}^{(1,1)}$                & 630.39 & $1.31$     & $b W_\mu^{(1,1)}$              & $100\%$  \\ 
\midrule
$d_{s}^{(1,1)}$        & 600.89 & $3.24$     & $d A_\mu^{(1,1)}$ & $99.8\%$ \\ 
                               &        &                         & $d Z_\mu^{(1,1)}$      & $0.2\%$  \\
\midrule
$b_{s}^{(1,1)}$        & 600.89 & $3.24$     & $b A_\mu^{(1,1)}$ & $99.8\%$ \\ 
                               &        &                         & $b Z_\mu^{(1,1)}$      & $0.2\%$  \\
\midrule
$u_{d}^{(1,1)}$          & 608.58 & $26.6$     & $u A_\mu^{(1,1)}$   & $13.5\%$ \\ 
                               &        &                         & $u Z_\mu^{(1,1)}$        & $26.8\%$ \\
                               &        &                         & $d W_\mu^{(1,1)}$        & $59.7\%$ \\
\midrule
$t_{d}^{(1,1)}$                & 637.01 & $87.8$     & $b W_\mu^{(1,1)}$              & $100\%$  \\ 
\midrule
$d_{d}^{(1,1)}$          & 608.58 & $23.8$     & $d A_\mu^{(1,1)}$   & $0.4\%$  \\ 
                               &        &                         & $d Z_\mu^{(1,1)}$        & $33\%$   \\
                               &        &                         & $u W_\mu^{(1,1)}$        & $66.6\%$ \\
\midrule
$b_{d}^{(1,1)}$                & 612.37 & $11.1$     & $b A_\mu^{(1,1)}$         & $0.9\%$  \\ 
                               &        &                         & $b Z_\mu^{(1,1)}$              & $99.1\%$   \\
\midrule
\midrule
\multicolumn{4}{c}{Leptons} \\
\midrule
\midrule
$l_{L}^{(1,1)}$    & $574.43$ & $0.65$     & $l    A_\mu^{(1,1)}$   & $100\%$ \\ 
\midrule
$l_{R}^{(1,1)}$    & $568.87$ & $0.16$     & $e    A_\mu^{(1,1)}$   & $100\%$ \\ 
\midrule
$\nu_l^{(1,1)}$& 574.43 & $0.24$     & $\nu_l A_\mu^{(1,1)}$   & $100\%$ \\ 
\bottomrule
\end{tabular}} \end{center}
\caption{Masses, widths and decay channels with branching ratios of states in tier $(1,1)$: here, $u$ stands for up and charm quarks, $d$ for down and strange, $l$ for electron, muon and tau leptons. The width of $A_\mu^{(1,1)}$ is estimated in Appendix~\ref{app:locops}.}
\label{tab:11spectrumdecaychannels}
\end{table}

\subsection{The four top final state}
\label{subsec:4top}

A state in the tier $(1,1)$ can be either pair produced via bulk interactions, whose size is similar to SM couplings, or be singly 
produced via localised interactions: in this work we are assuming that the single production is very small compared to the 
pair production as it is mediated by higher order operators.
Once a heavy state of the tier $(1,1)$ is produced, it will undergo chain decays until the lightest state, 
namely the vector photon $A_\mu^{(1,1)}$, is reached.
The SM particles radiated in the chain decay are soft due to the typically small mass differences between states in the tier, and therefore they will easily escape detection or be hidden in the large SM soft background.
Therefore, the production of any pair of $(1,1)$ states effectively contributes to the production of a pair of $A_\mu^{(1,1)}$ vectors.
This cumulative effect ensures that a large number of heavy vector photons will be produced at the LHC, even though their couplings to quarks and gluons are small.

The fate of the heavy photon depends on the localised interactions, that mediate its decays into a pair of SM particles.
Details about this final decay are crucial for subsequent analysis. 
A localised interaction is described by parameters which cannot be predicted within the effective
theory, but can only be determined by knowing the UV-completion of the theory.
Due to our ignorance of the details of such UV-completion, we can only give reasonable assumptions on the size of the coupling and on the branching ratios of the decay of the $(1,1)$ vector photon into SM states. 
Here we will focus on the decay into a pair of tops, $t \bar t$: the rationale under this assumption is that the top is the heaviest particle in the SM and therefore it may couple more strongly to the UV completion of the model.
Moreover, tops usually play a crucial role in the electroweak symmetry breaking sector of the model due to their large coupling to the Higgs, which is very sensitive to UV physics.
We will parametrise our ignorance of the UV physics as $b= BR (A_\mu^{(1,1)} \to t \bar{t})$, so that the cross section for the 4-top final state will be given by
\begin{equation}
\sigma_{\rm tot} (t \bar{t} t \bar{t})= \sigma_{\rm eff} (A_\mu^{(1,1)} A_\nu^{(1,1)}) \times b^2\,.
\end{equation}
In our simulation, we will assume $b=100\%$ to generate the signal; in case of a null result in the CMS search for 4-top final states, we can pose an upper bound on the value of $b$.
Recall that $\sigma_{eff} (A_\mu^{(1,1)} A_\mu^{(1,1)})$ is only determined by the mass of the new states ($\approx \sqrt{2} m_{KK}$) and SM couplings.
In our Monte Carlo implementation of the model, we introduced an effective coupling of the heavy photon with tops with electromagnetic strength, suppressed by a loop factor:
\begin{equation}
 g_{A_\mu^{(1,1)}t\bar t}=\frac{Q_te}{16\pi^2}\,,
 \end{equation}
with $Q_t$=2/3. 
The precise value of this coupling, however, is not important for our results as long as the decay in tops is prompt and 
does not leave displaced vertexes (we will not consider this possibility here, neither the possibility that the heavy photon decays 
outside the detector). 
The operator that mediates the decay can be found in Appendix~\ref{app:locops}, and it gives rise to a partial width of order $10^{-6}$ GeV.

As the heavy $(1,1)$ states can only be produced in pairs, after the subsequent chain decays, 
the final state will always contain two heavy vector photons which can finally decay into two $t\bar t$ pairs.
At the LHC the 4-top productions takes place as follows :
\begin{equation}
pp \to P^{(1,1)} {P'}^{(1,1)} \to A_\mu^{(1,1)} A_\mu^{(1,1)} + X \to t\bar t t\bar t +X
\end{equation}
where $X$ are the jets and the leptons coming from the decay chain of the $(1,1)$ states $P$ and $P'$.
All channels will lead to such final state; however, the main production cross section will involve coloured states like heavy gluon 
and quark resonances.

The total cross section is mainly accounted for by the following three basic processes that, in the benchmark point $m_{KK} = 400$ GeV,  amount to:
\begin{eqnarray}
 \left.\begin{array}{rcl}
 \sigma_{pp\to q^{(1,1)}q^{(1,1)}} &\simeq& 7.6~pb\\
 \sigma_{pp\to g_\mu^{(1,1)}q^{(1,1)}} &\simeq& 5.5~pb\\
 \sigma_{pp\to g_\mu^{(1,1)}g_\mu^{(1,1)}} &\simeq& 0.7~pb
 \end{array}\right\} \quad
 \sigma_{\rm eff}   (A_\mu^{(1,1)} A_\mu^{(1,1)}) \gtrsim 13~pb~@~7~TeV\,.
\label{eq:xsect}
\end{eqnarray}
Other processes involving heavy electroweak gauge bosons and leptons can be neglected; from now on we shall also neglect
the heavy gluon pair production process and processes involving the heavier top states.
The decay channels of intermediate states, together with their BR's, are shown in Tab.\ref{tab:11spectrumdecaychannels}.
The relevant decays for the processes in Eq.~\ref{eq:xsect} are:
\begin{eqnarray}
\begin{array}{rclll}
 q_{s,d}^{(1,1)}  \to & & q~A_\mu^{(1,1)} & & \longrightarrow t\bar t + 1~\textrm{jet} \\
 q_d^{(1,1)}      \to &  q~\{Z,W\}_\mu^{(1,1)} \to & q~l~l^{(1,1)} & \to q~2l~A_\mu^{(1,1)} & \longrightarrow t\bar t+ 1~\textrm{jet} + 2l \\
 g_\mu^{(1,1)}    \to & & q~q_{s,d}^{(1,1)} & & \longrightarrow \left\{\begin{array}{l} t\bar t + 2~\textrm{jets} \\  t\bar t + 2~\textrm{jets} + 2l\end{array}\right.
\end{array}
\label{eq:finalstates}
\end{eqnarray}
where leptons include neutrinos (therefore the lepton pairs can be $l^+ l^-$, $\nu_l \bar{\nu}_l$, $l^+ \nu_l$ and $l^- \bar{\nu}_l$, where $l=e,\mu,\tau$).
The final state will contain all possible combinations of the above:
\begin{equation}
 pp \to t\bar tt\bar t + n~\textrm{light jets} + n'~\textrm{leptons} 
\end{equation}
with $n=2,3,4$ and $n'=0,2,4$.
Noticeably, jets and leptons produced together with the final 4 tops are mostly soft since
the mass splittings between $(1,1)$ particles are few tens of GeV at most. 
The only way to make such soft particles detected is to produce the heavy states with a large transverse momentum, however 
the cross section would be largely suppressed. The final state, therefore, mainly consists of the products of the four top decays,
 which carry enough momentum from the decay of the heavy photon. 
In the following we shall simulate part of the signal in the 6D model including the soft SM products of the chain decay. 

\begin{figure}
\centering\epsfig{file=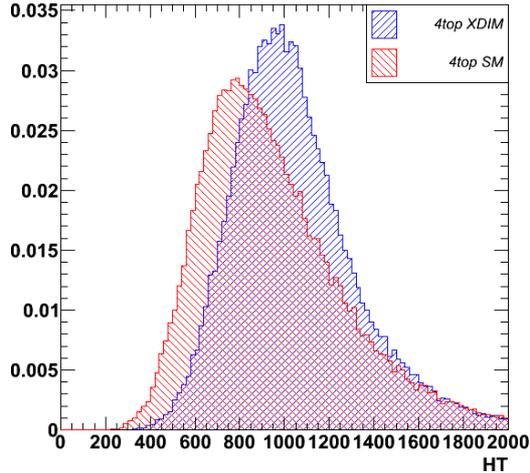, width=0.45\textwidth, angle=0}
\caption{Comparison of differential cross sections normalised to unity with respect to HT (scalar sum of the $p_T$ of jets) for 
the production  of $t\bar t t \bar t$ in the RPP model (benchmark point with $M_{KK} = 400$ GeV) and in the SM.
These events include the effect of hadronisation and detector efficiencies (for more details, refer to Section~\ref{subsec:detsimu}.}
\label{smZvsxd}
\end{figure}

It is useful to note at this point that the softness of the SM debris of the chain decay may allow us to use a simplified model to study this phenomenology: one can generate the same signal in a model with a massive vector $Z'$ that couples to tops.
The mass of the $Z'$ plays the role of the mass of the tier, $\sqrt{2} M_{KK}$, while the normalisation of the cross section can be rescaled to match the production cross section of heavy quarks and gluons.
By use of an implementation of both models in  \verb#Madgraph4/Madevent#\cite{Alwall:2007st}, we checked that, at parton level, the kinematic distribution are the same to a good approximation, proving that the two simulations would give rise to the same bounds on $M_{KK}$.
Here, however, we did not make use of this simplification.
Another more rough possibility is to rescale the SM signal, which is given mainly by QDC diagrams: in this case, there is no information about the mass scale of the new particle, and one can just obtain a conservative bound on the total cross section.
In Figure~\ref{smZvsxd}, we show the distribution in HT of the events in the SM and in our benchmark point of the model.
We can clearly see that the SM tends to be softer, therefore a cut in HT used to reduce the background may underestimate the number of events.

\subsection{Four tops from tiers $(2,0)$ and $(0,2)$.}
\label{sec:pheno2}

The tiers $(2,0)$ and $(0,2)$ are quite interesting for phenomenology because they contain the 
lightest even states that can decay into SM particles via loop interactions without missing energy. 
The main difference with respect to the tier $(1,1)$, besides the higher mass, is that the decay rates can be calculated and 
predicted, with only a mild logarithmic dependence on the cut-off of the theory.
For instance, the lightest state in the tiers, which is a vector photon, decays into a pair of tops with a rate of about $20$\%.
This tiers, therefore, will also generate a 4 top final state that can be analysed in the same way as the one presented in this work.
The decay chains are typically more complex, because each state in the levels can decay directly into a pair of SM particles or 
into a pair of states in the odd levels $(1,0)$ or $(0,1)$ via bulk interactions.
All such rates can be calculated and predicted.
In~\cite{Cacciapaglia:2011hx} a more detailed study of the $(2,0)$ and $(0,2)$ decay channels has been presented.
The main production mechanism is pair production of coloured states, heavy quarks and gluons, which can in some cases 
chain decay into a final state of 4 tops (each pair of top-antitop coming from the decay of the vector photon, 
$A_\mu^{(2,0) - (0,2)}$ and, less frequently, from the vector $Z_\mu^{(2,0) - (0,2)}$).
The authors considered a benchmark point with $m_{KK} = 300$ GeV, for which the mass of the tier is around $600$ GeV 
similar to the mass of the $(1,1)$ state under consideration here.
An estimate of the inclusive 4-top cross section leads to about $77$ fb for each tier: this number is much smaller than the 
$(1,1)$ case because of the suppressed branching ratios into tops.
However, it is important to notice that this cross section cannot be suppressed by any arbitrary parameter in the theory, therefore 
a non observation of the 4-top signal at this level would lead to an exclusion of the model at this mass scale.

In the following, we will focus uniquely on the signal generated by the $(1,1)$ level, but keeping in mind that the same result can 
be applied to the smaller but irremovable signal from the $(2,0)$ and $(0,2)$ levels.

\section{Simulation details}
\label{sec:simul}
\setcounter{equation}{0}
\setcounter{footnote}{0}

In this section we will discuss some details of our simulation.
We will be interested in a beam energy of $7$ TeV in the centre of mass, and we will normalise the event number to an integrated luminosity of $1$ fb$^{-1}$, which has already been collected before the summer 2011 at the LHC.

\subsection{Signal}
\label{subsec:signal}

The UED/RPP model has been implemented in \verb#FeynRules#\cite{Christensen:2008py} to obtain the necessary Feynman rules for (1,1) couplings  
and the simulation has been performed exploiting standard software tools: 
diagram topologies and event simulations have been generated with \verb#Madgraph4/Madevent#\cite{Alwall:2007st}
for each possible final state containing 4 tops. The subsequent top decay and hadronisation has been obtained using \verb#Pythia6.4#\cite{Sjostrand:2006za}. \\

The relative relevance of different signatures can be computed considering the branching ratios of intermediate heavy particles decays.
Referring to Tab.\ref{tab:11spectrumdecaychannels} the total cross section is composed of:
\begin{eqnarray}
\begin{array}{rcl}
 \sigma_{pp \to 4t + 2j} \simeq 2.4~pb &\sim& 17\%\,, \\
 \sigma_{pp \to 4t + 2j + 2l} \simeq 3.1~pb &\sim& 22\%\,, \\
 \sigma_{pp \to 4t + 2j + 4l} \simeq 2.1~pb &\sim& 15\%\,, \\
\end{array}
\quad
\begin{array}{rcl}
 \sigma_{pp \to 4t + 3j} \simeq 1.9~pb &\sim& 14\%\,, \\
 \sigma_{pp \to 4t + 3j + 2l} \simeq 2.5~pb &\sim& 18\%\,, \\
 \sigma_{pp \to 4t + 3j + 4l} \simeq 1.1~pb &\sim& 8\%\,. \\
\end{array}
\label{weightsxsect}
\end{eqnarray}
Unfortunately, the large number of external particles makes it difficult to generate events 
for all possible final states with the available software: 
we therefore provide results of numerical simulations for the simplest signatures 
and extrapolate the rest of the signal, when possible.
Following (\ref{eq:xsect}), we first select signatures coming from the $q^{(1,1)}q^{(1,1)}$ intermediate state, and looking at (\ref{eq:finalstates})
it is possible to see that the most relevant final states are $4t + 2j + n'l$ with $n'=0,2,4$. 
Diagrams for a signal without leptons contain 8 external particles (including the initial partons) 
and this is the practical upper limit for our computation.
Signatures with more than 2 soft jets necessarily involve the generation of a heavy gluon, 
but even if in principle the signal $4t + 3~\textrm{soft jets}$ can be computed by Madgraph, 
in practice the combinatorial is huge and makes the computation unfeasible.
As already stressed, the 4 jet signal can be discarded since its cross section is one order of magnitude smaller than the rest.

We already noticed that jets and leptons emerging from intermediate decays of heavy states are mostly soft, 
thus signatures with different number of jets and leptons have very similar distributions. 
This means that it is sufficient to base our study on the kinematic details of the 4 tops in the final state, coming from the decays of the heavy photons: we have verified the validity of this assumptions by comparing the events with two soft jets with events with one or no jet (that can be easily generated), coming from sub-leading diagrams involving direct production of the heavy photon.
In the event generation, we have imposed minimal kinematic cuts on the transverse momentum and rapidity of light- and b-jets, namely 
$p_T >20$GeV and $\eta < 5$, 
while no cuts have been imposed on leptons. The PDFs used for the computations are the CTEQ6L1\cite{Pumplin:2002vw}.

\subsection{Identification of SM backgrounds}
\label{subsec:smbg}
Reconstructing the four top quarks individually is a very difficult task, which suffers from huge combinatorial background.
In addition, in the complex environment of the LHC, final state objects overlap and their individual reconstruction is not efficient.
In order to get rid of the huge multijet background, we study the potential of a same sign dilepton signature. This channel accounts for 21\%  (at tree level) of the signal.

In the following we consider only electrons and muons. 
There are two main categories of backgrounds: one is represented by $t\bar{t}$+jets, $V$qq (where $V$=$W$ or $Z$), 
$VV$+jets and $Z$+jets. They do not contain opposite sign dileptons, but their cross-section is large enough to fake a same 
sign dilepton signature because of the charge misidentification of the leptons. 
The second category of background is the one with two real same sign dileptons. We have considered 
$Zt\bar{t}$+jets and $Wt\bar{t}$+jets and the $t\bar{t}t\bar{t}$ production in the SM.
The cross section of the aforementioned processes is smaller, but they yield an irreducible background.
All these backgrounds and their cross-sections are summarised in table \ref{tab:bckgdescription}.

\begin{table}[t!]
\begin{center}
\begin{tabular}{cc}
\toprule
Background process & cross-section [pb]\\
\midrule
$Z$+jets          &   2800 \\
$t\bar{t}$+jets &  165    \\
$Vqq$              & 36       \\ 
$VV$+jets       &   5        \\
$Wt\bar{t}$      & 0.018  \\
$Zt\bar{t}$       &  0.013 \\
SM  $t\bar{t}t\bar{t}$ & 0.010 \\
\bottomrule     
\end{tabular}  
\caption{List of backgrounds considered in this analysis. The cross sections for all processes are at the LO, as 
obtained from the {\tt MadEvent} calculator~\cite{Alwall:2007st}, except for $t\bar{t}$ events for which the approximate 
NLO calculation of~\cite{Kidonakis:2009mx} is used. }
\label{tab:bckgdescription}
\end{center}
\end{table}

\subsection{Detector simulation}
\label{subsec:detsimu}

As explained before, background and signal events are generated with {\tt MadGraph/MadEvent}. In order to take into account the efficiency of event selection under realistic experimental conditions, the detector simulation is carried out with the fast detector simulator {\tt Delphes}~\cite{Delphes} using the CMS detector model. The tracker is assumed to reconstruct tracks within $|\eta | < 2.4$ with a 100\% efficiency and the calorimeters cover a pseudo-rapidity region up to  $|\eta | <$ 3 with an electromagnetic and hadronic tower segmentation corresponding the the CMS detector. The energy of each quasi stable particle is summed up in the corresponding calorimeter tower. The resulting energy is then smeared according to resolution functions assigned to the electromagnetic calorimeter (EC) and the hadronic calorimeter (HC). The acceptance criteria are summarised in the Tab.~\ref{Tab:DelphesAcceptance}. 
For the leptons (electrons of muons), we require a tight isolation criterion: no additional tracks with $p_T> 2$ GeV must be 
present in a cone $\Delta R = \sqrt{\Delta \eta^2 + \Delta \phi^2}= 0.5$ centred on the lepton track. The jets are reconstructed using only the calorimeter towers, through the anti-kt algorithm with cone size radius of 0.5, as defined in the {\tt FastJet} 
package~\cite{fastjet}  and implemented in {\tt Delphes}. The b-tagging efficiency is assumed to be 40\% for all b-jets, independently of their transverse momentum, with a fake rate of 1\% (10\%) for light (charm) jets. Finally, the total missing transverse energy (MET) is reconstructed using information from the calorimetric towers and muon candidates only.

In the following, we present a simple strategy that can lead to promising Signal-over-Background (S/B) ratios. Our purpose is to illustrate the new possibilities that open up in the four top final state and motivate more detailed studies. To this aim detailed information on the efficiencies and the visible cross sections are given. 

\begin{table}[t!]
\begin{center}
\begin{tabular}{lcc}
\toprule
 object & $|\eta_{max}|$ & $p_T^{min}$ [GeV]\\
\midrule
e,$\mu$ & 2.4 & 10 \\ 
jets & 3 & 20 \\
b-jets & 2.5 & 20 \\
\bottomrule     
\end{tabular}  
\caption{Acceptance of the different final states in the simulated detector.}
\label{Tab:DelphesAcceptance}
\end{center}
\end{table}


\begin{figure}
\centering
\subfloat[]
{\epsfig{file=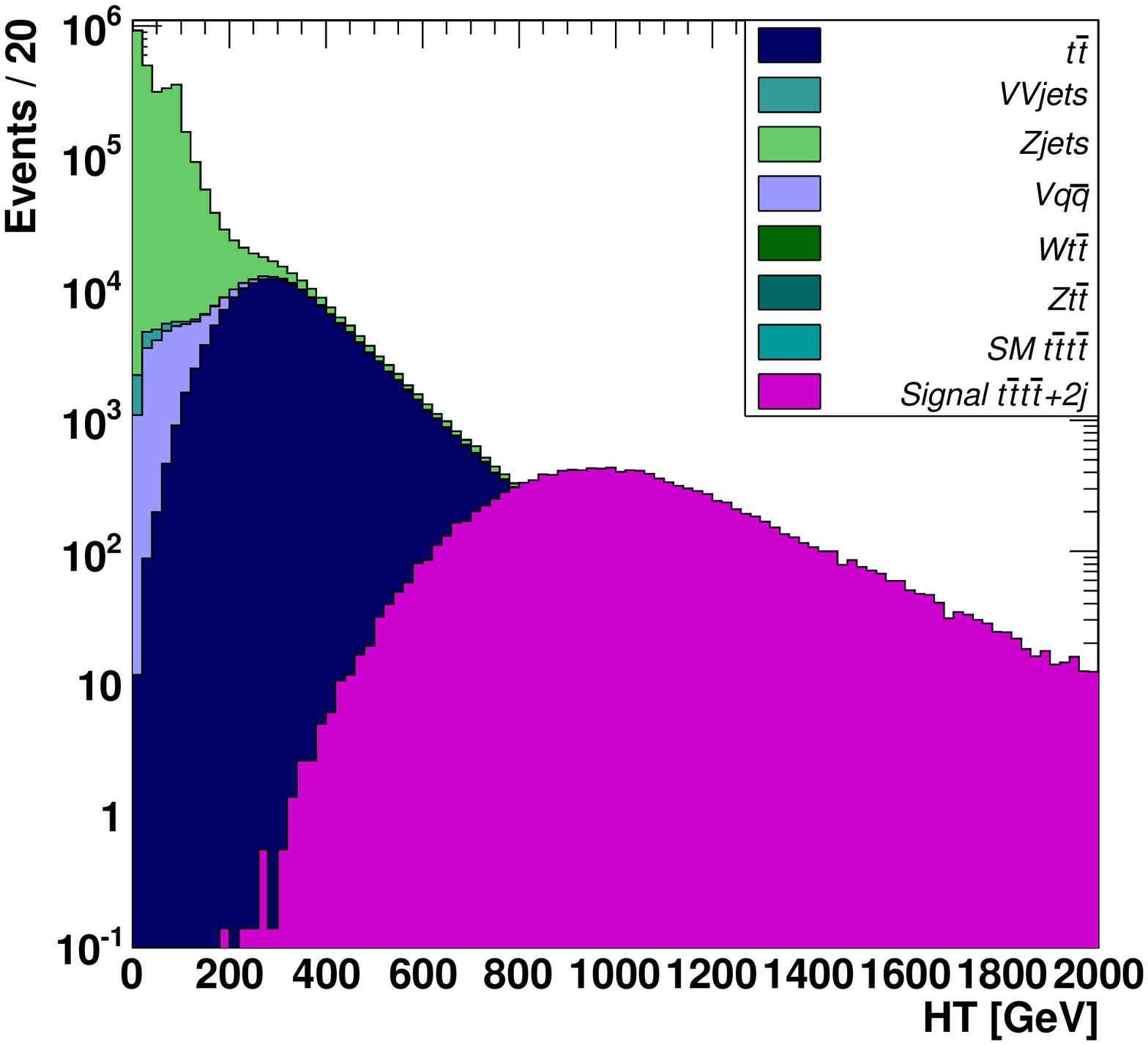, width=0.45\textwidth, angle=0}}\hfill
\subfloat[]
{\epsfig{file=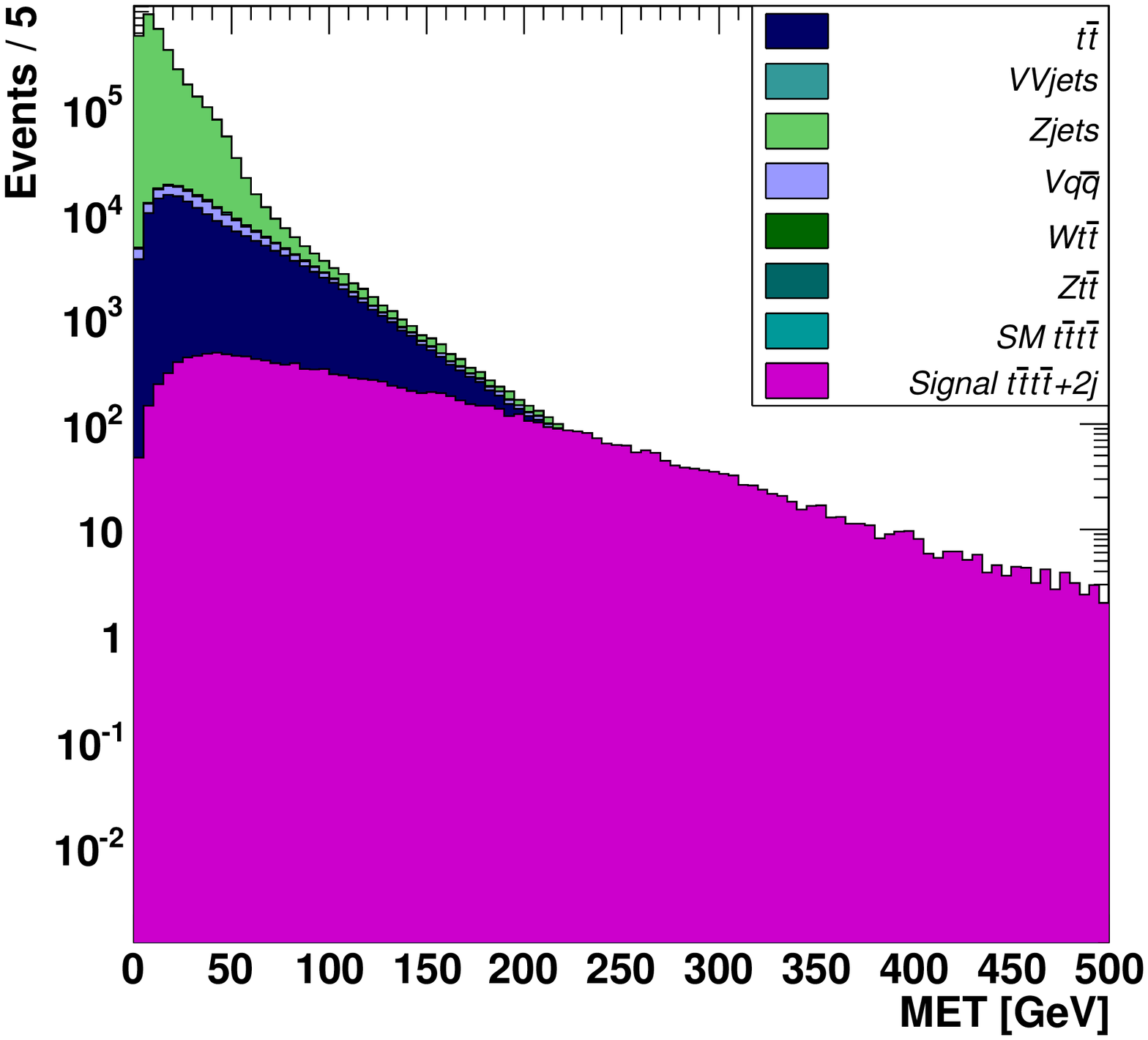, width=0.45\textwidth, angle=0}}\\
\subfloat[]
{\epsfig{file=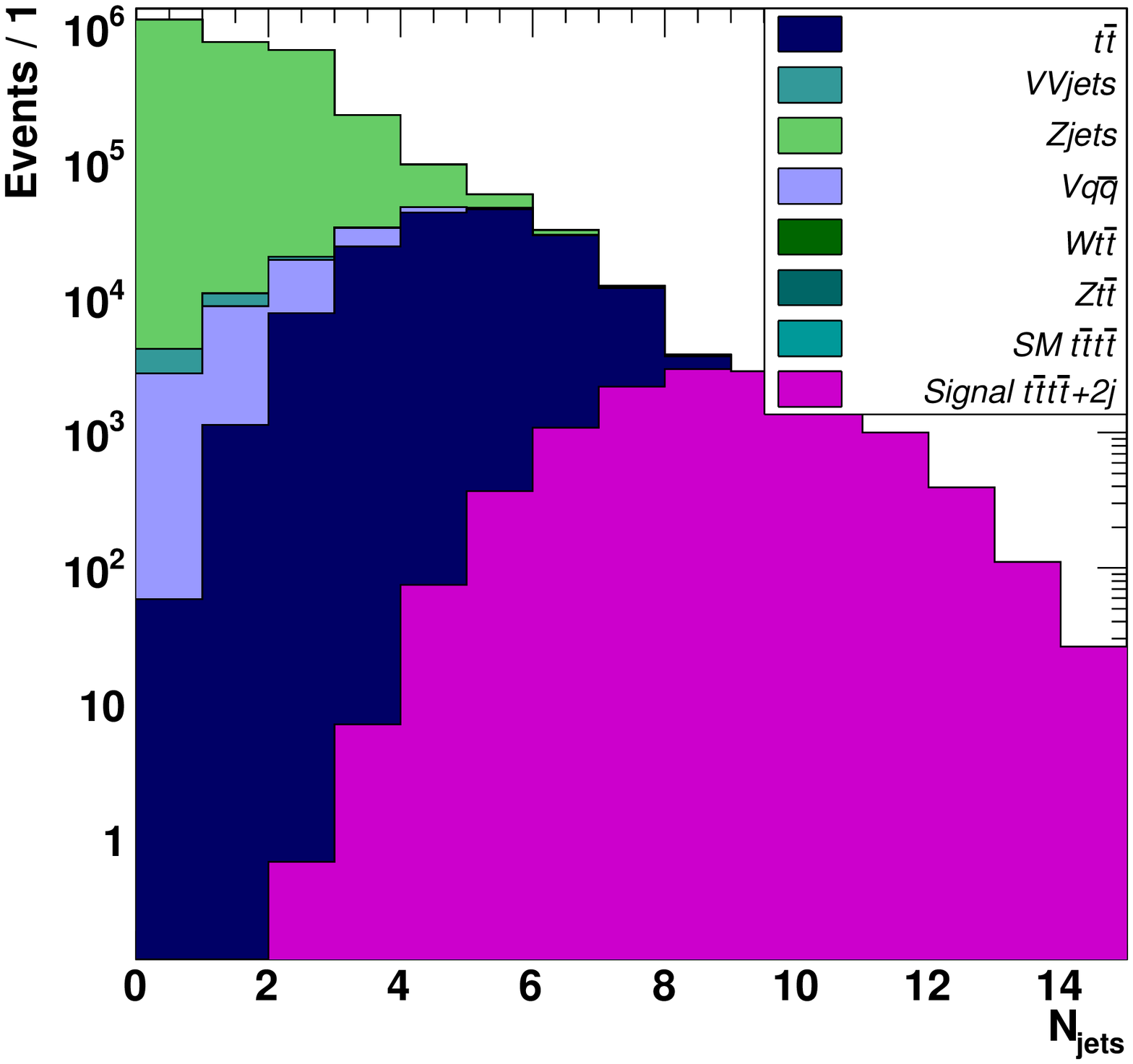, width=0.45\textwidth, angle=0}}\hfill
\subfloat[]
{\epsfig{file=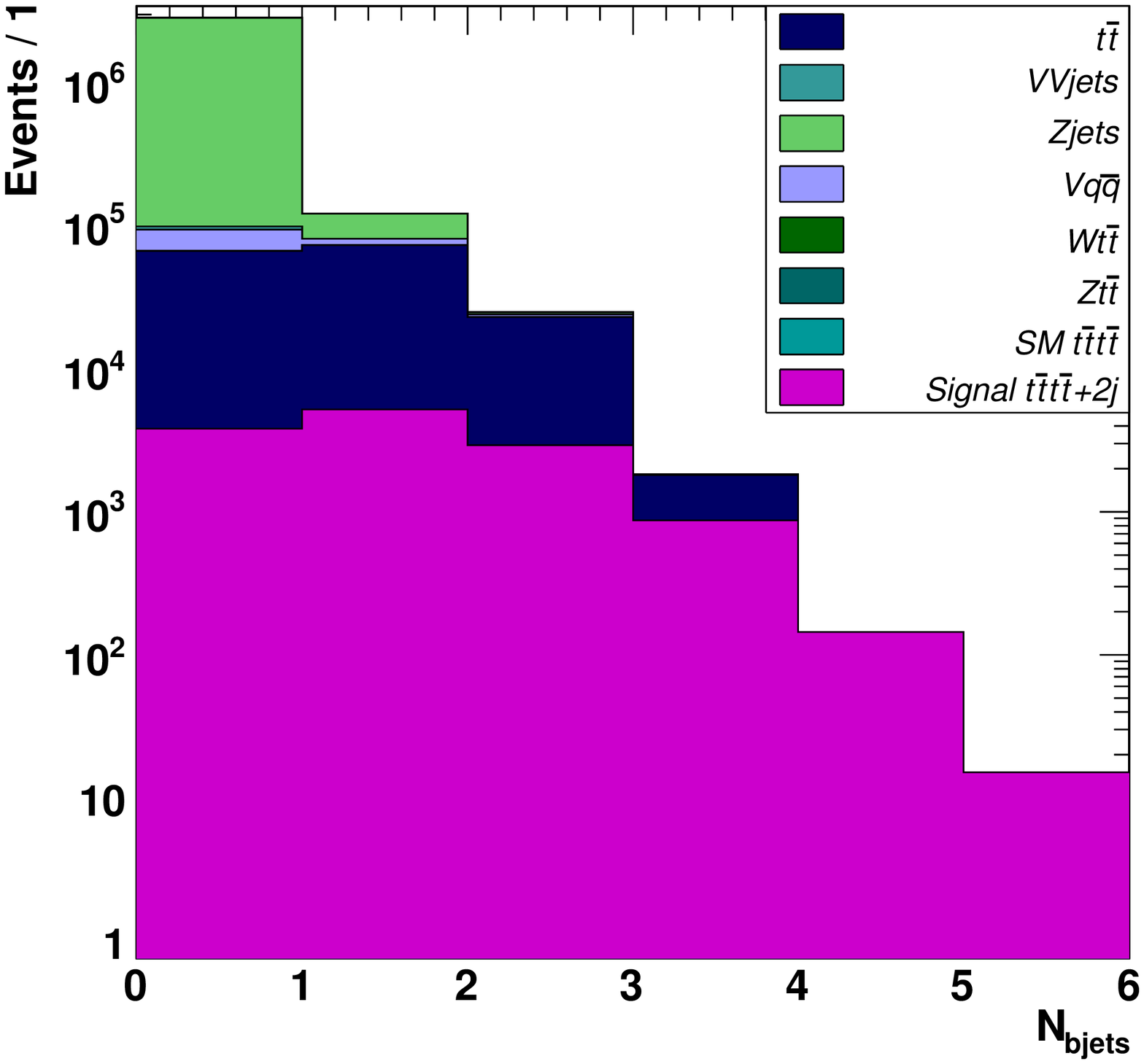, width=0.45\textwidth, angle=0}}
{\caption{Differential cross sections for the signal and SM backgrounds (stacked histogram) without any selection cut with respect to (a) HT (the scalar sum of the $p_T$ of the jets), (b) MET (missing transverse energy), (c) the number of hard jets and (d) the number of b-tagged jets. The $b$ parameter is set to one in order to allow an easy rescaling of the distributions.} 
\label{fig:distnosel}}
\end{figure}

\begin{figure}
\centering
\subfloat[]
{\epsfig{file=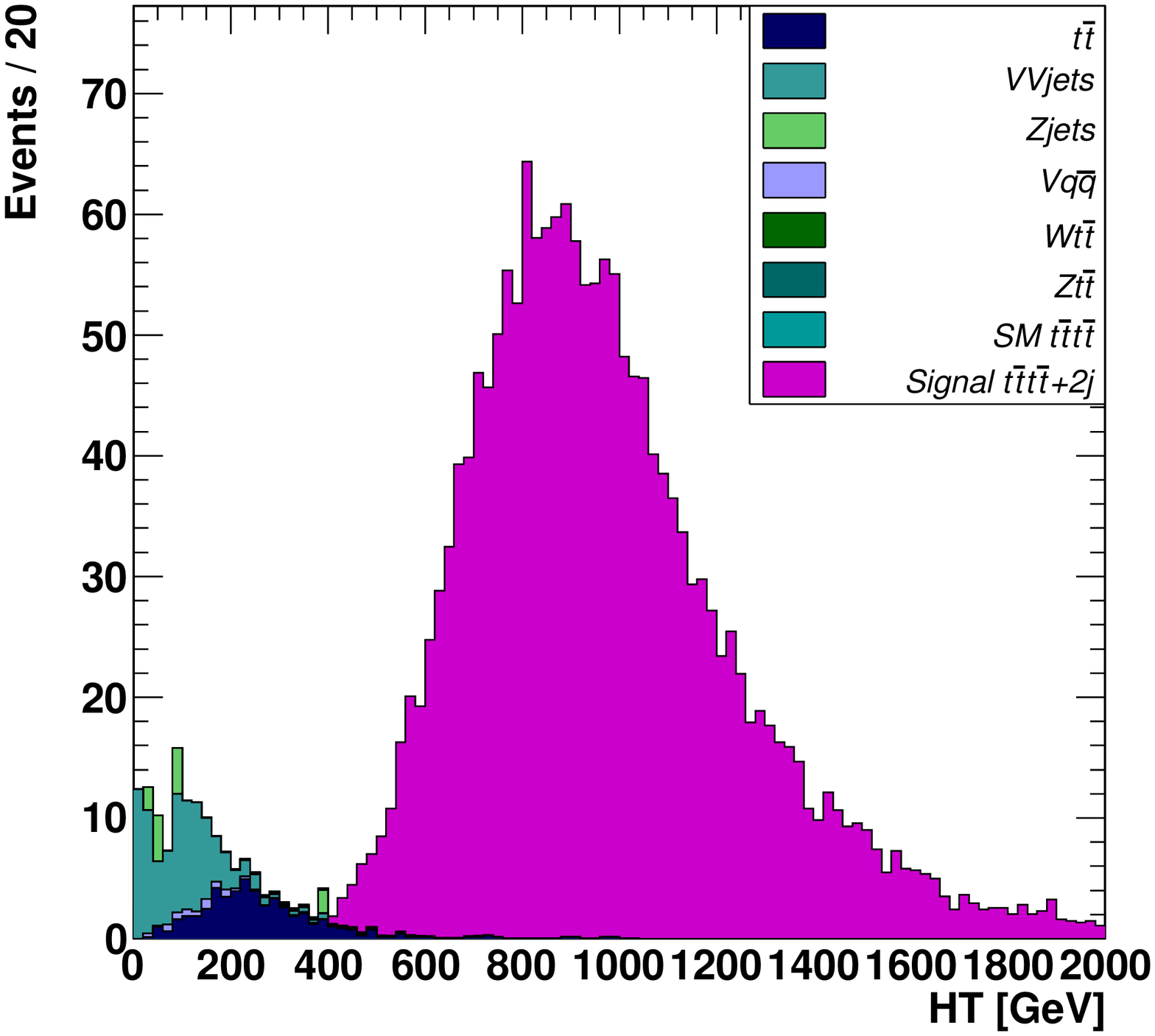, width=0.45\textwidth, angle=0}}\hfill
\subfloat[]
{\epsfig{file=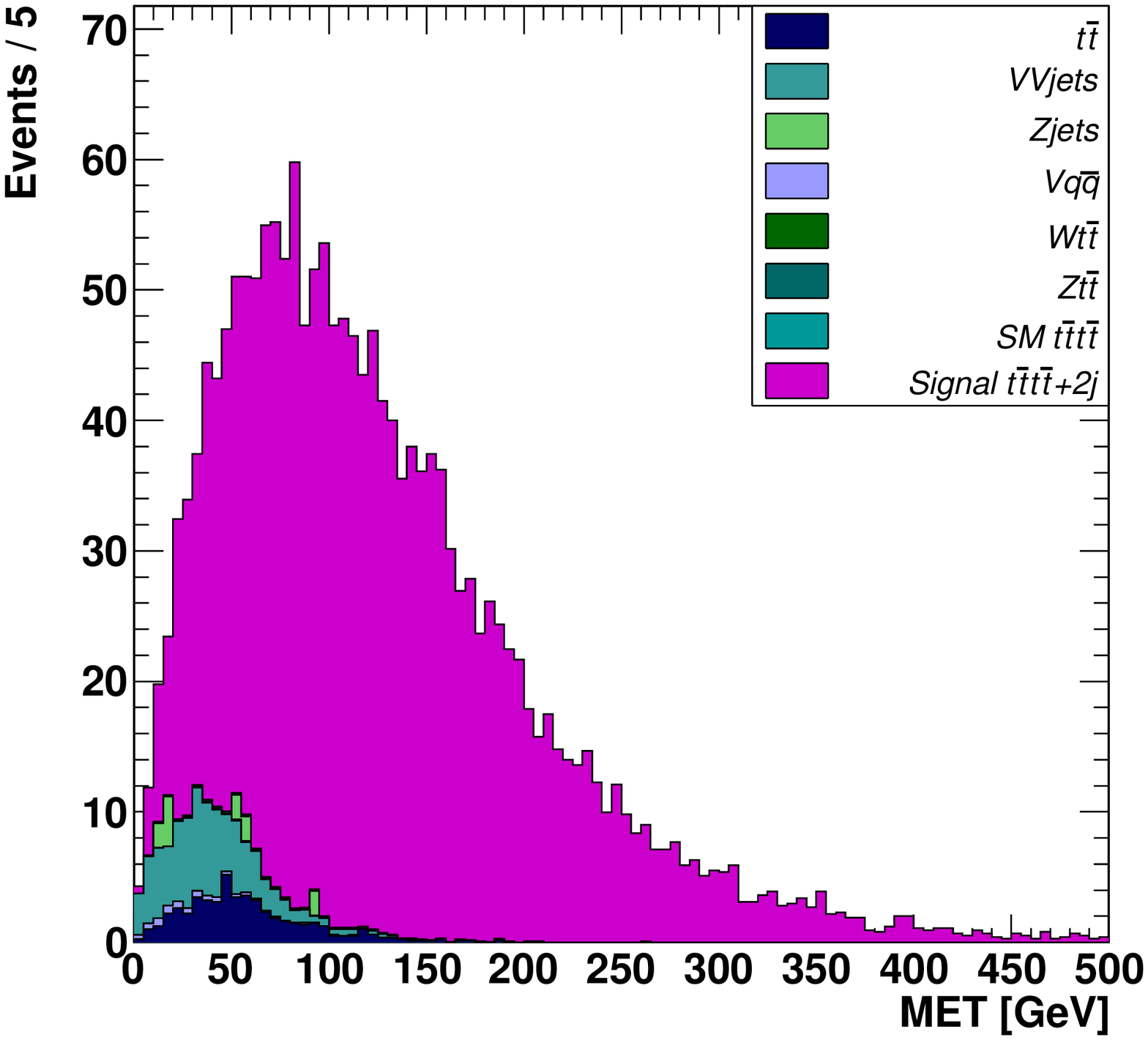, width=0.45\textwidth, angle=0}}\\
\subfloat[]
{\epsfig{file=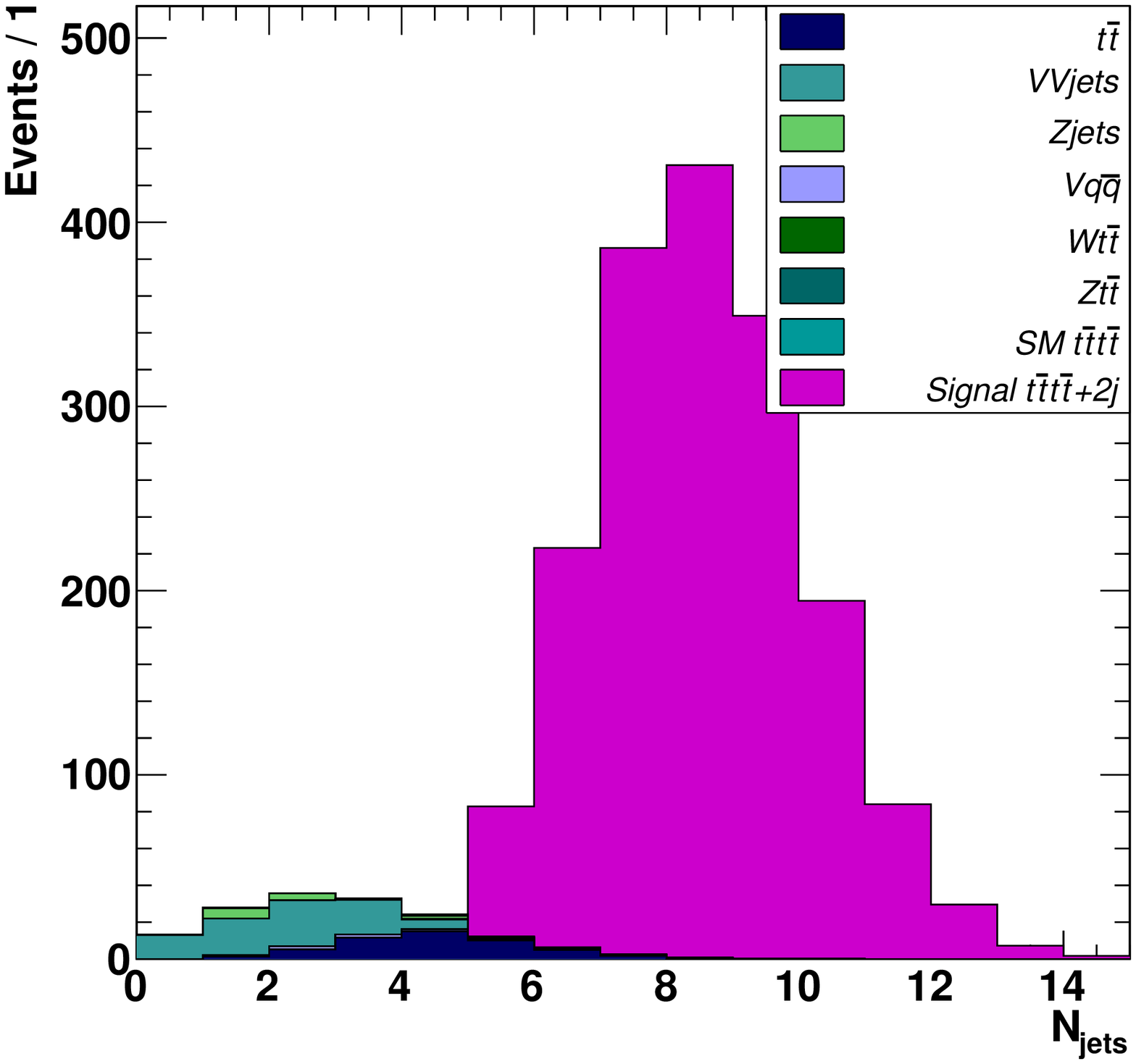, width=0.45\textwidth, angle=0}}\hfill
\subfloat[]
{\epsfig{file=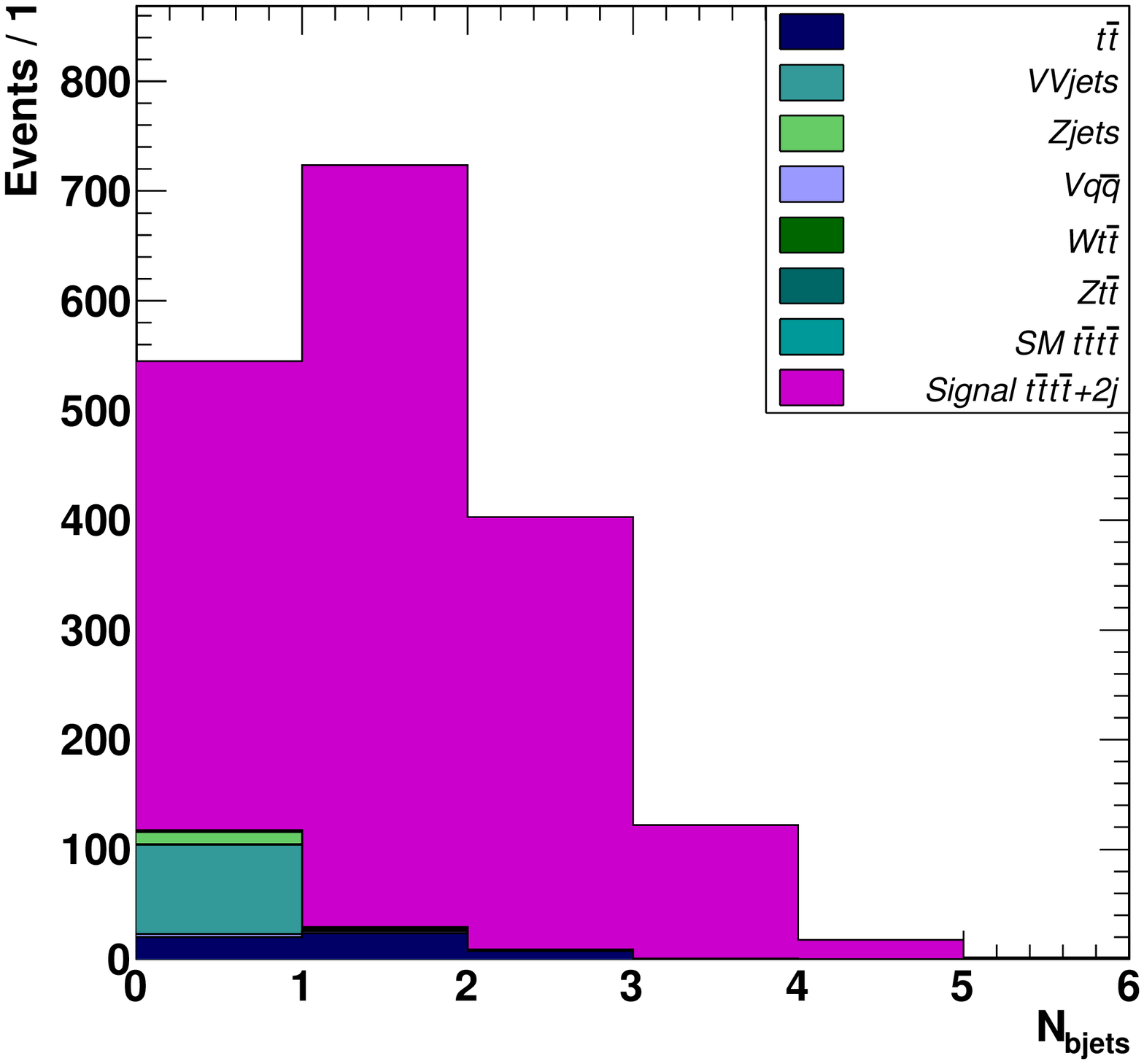, width=0.45\textwidth, angle=0}}
{\caption{Differential cross sections for the signal and SM backgrounds (stacked histogram) after the requirement of two same sign leptons with respect to (a) HT, (b) MET, (c) the number of hard jets and (d) the number of b-tagged 
jets. The $b$ parameter is set to one in order to allow an easy rescaling of the distributions.} 
\label{fig:distlept}}
\end{figure}

\subsection{Analysis results}
\label{subsec:detsimu}
\begin{figure}
\centering
{\epsfig{file=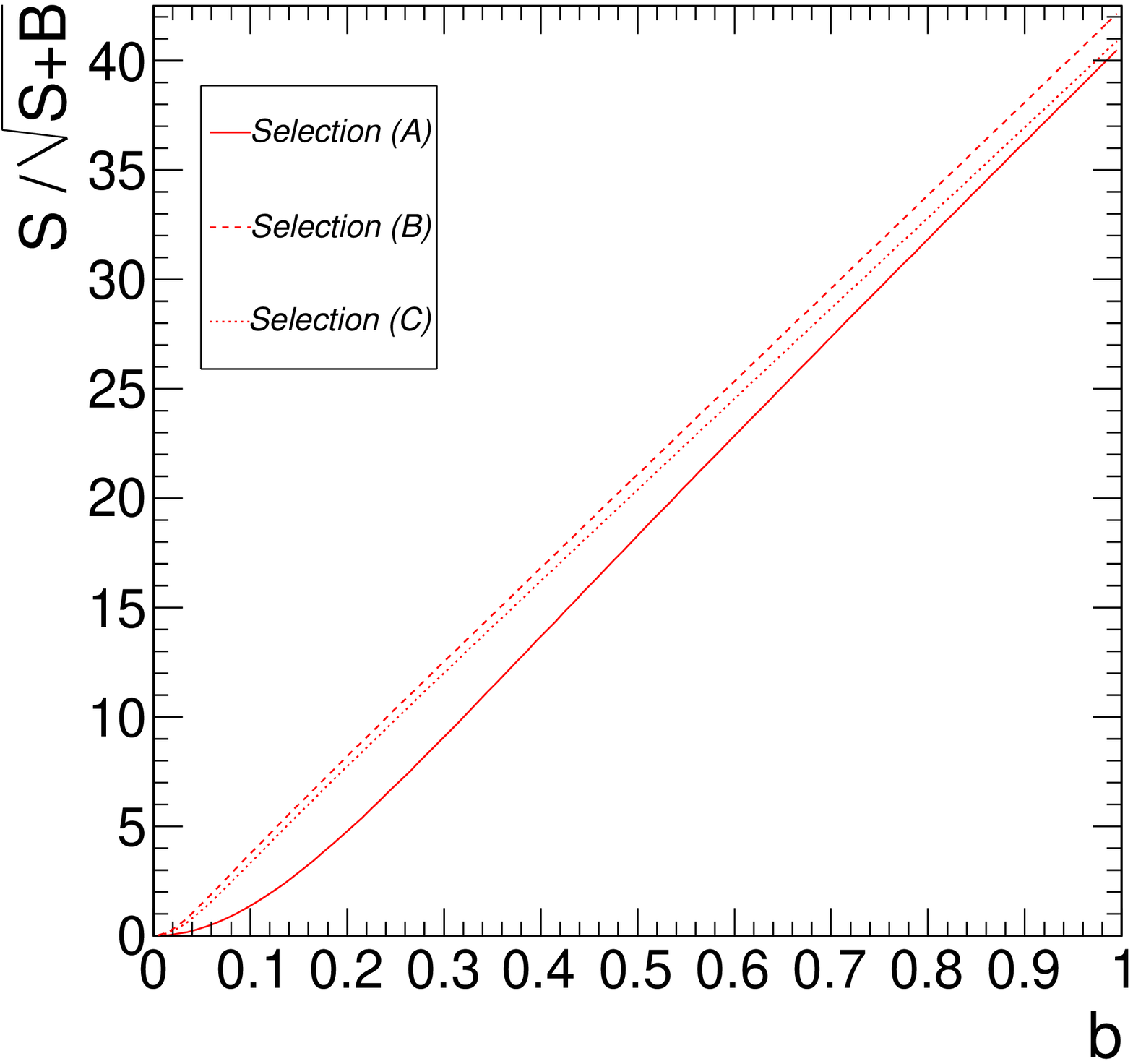, width=0.45\textwidth, angle=0}} {\epsfig{file=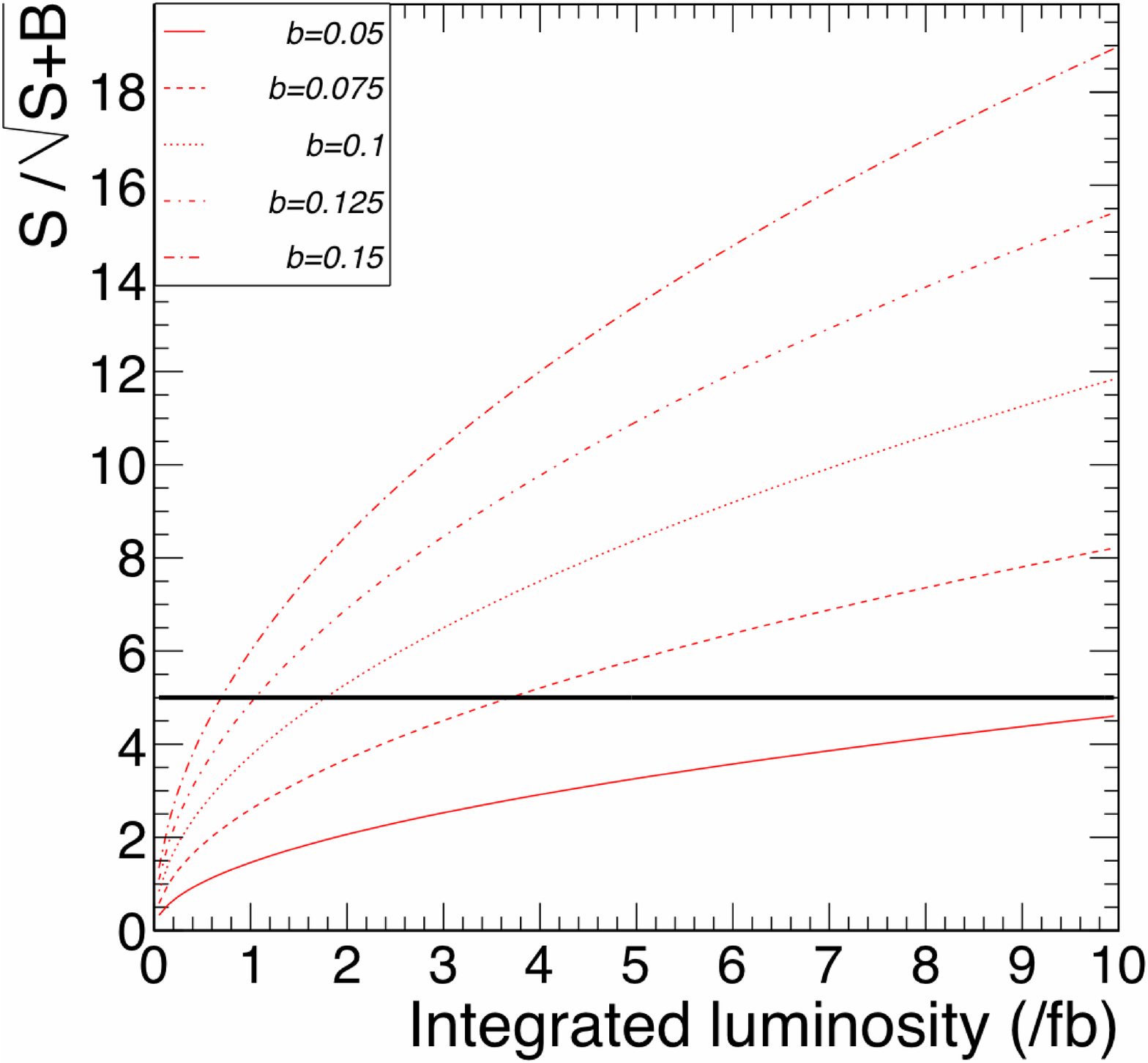, width=0.45\textwidth, angle=0}}\hfill
{\caption{Left: Evolution of the significance ratio $S/\sqrt{S+B}$ for the three selections with respect to the branching ratio parameter $b$, showing that selection (B) is the most effective.
Right: Evolution of the significance ratio $S/\sqrt{S+B}$ for the selection (B) and for different values of the branching ratio $b$ with respect to the integrated luminosity. The line indicating a significance ration of 5 needed for a discovery is also provided.} 
\label{fig:significance}}
\end{figure}

As outlined before, we request two isolated same sign leptons in the final state. Two well reconstructed
leptons with $p_T>30$ GeV and $|\eta | < 2.4$ are sufficient to trigger the events (the two lepton combined trigger efficiency 
is assumed to be 100\% after the lepton selection cut). Selected jets have $p_T>$20 GeV and $|\eta |<$ 3. 
The distributions for various observables without any selection cuts are are shown in Fig.~\ref{fig:distnosel} where the total signal cross-section is normalised to 13 $pb$, assuming a branching ratio $b=1$.
As expected, since the signal contains four tops, the distributions of signal and background events are well 
separated for HT (which is defined as the scalar sum of all the selected jets), and the number of jets. However this effect 
is less striking for MET, since some of the SM backgrounds have the same number neutrinos as the signal and for the number of b-jets.
Fig.~\ref{fig:distlept} shows the same distributions after the requirement of the same sign isolated leptons and a veto on the dilepton mass around the Z peak $\pm$ 10 GeV (to remove the Z+jets component).
With two like sign leptons, a four top final state produces eight hard quarks (including the four bottom quarks). QCD radiation can increase this further. 
The number of jets is around four for the the SM backgrounds with the $t\bar{t}$ component around 6 jets, while it peaks around 
nine jets for the signal (the dominant signal distribution comes from the four tops plus two extra jets contribution). The four top 
quarks also produce four bottom quarks when they decay. However the probability of tagging all four bottoms is rather small. 
Fig.~\ref{fig:distlept} shows the number of b quarks identified by Delphes. The four top signal peaks around 1 b being reconstructed, 
whereas the background is almost exclusively present in the 0 b-tagged jet bin. This variable is not optimal to discriminate 
between signal and background but it could help to verify if four tops are involved. However, the variable providing the most 
striking separation between SM backgrounds and signal is HT. Fig.~\ref{fig:distlept} shows that a cut around 450 GeV removes
almost all the remaining background and retains roughly all the four tops events. 
In the end, we chose to investigate the following selections: two like sign isolated leptons with $p_T>30$ GeV (selection A), a cut at 450 GeV on HT (selection B) and a number of jets greater or equal to 6 (selection C). 
Table~\ref{Tab:yield} presents the background and signal event yield, the Signal over noise ratio and the significance 
$S/\sqrt{S+B}$ for an integrated luminosity of 1/fb and a branching ratio $b=1$. The different selections are comparable 
with a final significance around 40. 
The left panel in Fig.~\ref{fig:significance} shows the evolution of the significance ratio $b^2 S/\sqrt{b^2 S+B}$ as a function of the branching ratio 
parameter $b$ for an integrated luminosity of 1/fb. For selection (B) a significance of 5 is obtained for $b=0.14$.
In order to claim a discovery, the significance ratio must be greater than 5. 
The right panel in Fig.~\ref{fig:significance} shows the evolution of the significance ratio for the selection (B)
and for different values of $b$ as a function of the integrated luminosity. For $b < 0.05$ no discovery is possible for an integrated luminosity below $10$ fb$^{-1}$.
For some larger values of $b$, the luminosity needed for a discovery is shown in Table~\ref{Tab:discovery}.

\begin{table}[t!]
\begin{center}
\begin{tabular}{lcccc}
\toprule
 Selection & Signal yield & Background yield & $S/B$ & $S/\sqrt{S+B}$ \\
\midrule
(A) & 1799 & 155 &   11.6 & 41 \\
(B) & 1786 & 5     & 357 & 42  \\
(C) & 1698 & 9     & 189 & 41  \\
\bottomrule     
\end{tabular}  
\caption{Background and signal event rates for integrated Luminosity of $1$/fb and for a branching fraction $b=1$, after various selection cuts. Selection (A) requires two like sign isolated leptons; selection (B) two like sign isolated leptons plus $HT>450$ GeV; finally selection (C) two like sign isolated leptons plus at least 6 jets. $S/B$ and $S/\sqrt{S+B}$ ratios are also provided for each selection.}
\label{Tab:yield}
\end{center}
\end{table}

\begin{table}[t!]
\begin{center}
\begin{tabular}{lcccc}
\toprule
 $b$ & Luminosity (/fb)\\
\midrule
0.05     & $>$ 10 \\
0.075  & 3.73 \\
0.1      & 1.76 \\
0.125 & 1.05 \\
0.15   & 0.70 \\
\bottomrule     
\end{tabular}  
\caption{Integrated luminosity needed for a discovery for different values of the branching fraction parameter $b$. }
\label{Tab:discovery}
\end{center}
\end{table}



\section{Conclusions}
\label{sec:conclusion}

We have studied the four top final state $t{\bar t}t{\bar t}$ at the 7 TeV Large Hadron Collider as a probe of physics beyond the 
standard model. The enhancement of the corresponding cross-section with respect to the Standard Model value allows to test 
and constrain extra dimensional models with heavy Kaluza-Klein gluons and quarks. In particular we have performed a 
detailed analysis including background and detector simulation for a universal extra--dimensional model with two extra 
dimensions compactified on a flat real projective plane. 
This background is especially interesting because it has an exact parity that forbids the decay of the lightest KK state, therefore offering a good candidate for Dark Matter.
Production cross-sections into $t{\bar t}t{\bar t}$ final states may be large for the 
$(1,1)$ tier of the model. 
When any heavy state of the tier $(1,1)$ is produced, it undergoes chain decays to the lightest state, 
the vector photon $A_\mu^{(1,1)}$, via bulk interactions. 
Finally the vector photon decays directly into SM particles through suppressed localised interactions.
If the heavy vector decays into tops, all the production channels participate in the 4-top final state, thus enhancing significantly the effective cross-section.
The size of the coupling and the branching ratio of the decay of the $(1,1)$ vector photon to SM states cannot be determined 
in the effective model, however it is constrained by LHC data. 
On the other hand, the heavy gauge bosons of the tiers $(2,0)$ and $(0,2)$, which are slightly heavier than the $(1,1)$, can decay into a pair of energetic SM fermions via calculable loops. 
Even though the branching ratios are rather small, due to the presence of other channels, and give rise to a smaller effective
cross-section in the four top final state, however in this case the coupling is 
determined in the effective theory in terms of SM parameters and the final cross-section can be computed.
Therefore, this signal can be a smoking gun or an exclusion arena for these models.
The background and signal events were generated with {\tt MadGraph/MadEvent}. The efficiency of event selection under realistic 
experimental conditions was simulated with the fast detector simulator {\tt Delphes} using the CMS detector model.
Reconstructing the four top quarks individually is very difficult and, in the complex environment of the LHC, final state 
objects overlap and their individual reconstruction is not efficient. 
We showed that selecting events with two same sign leptons from the 
$t{\bar t}t{\bar t}$ decays, where lepton means electron or muon, dramatically reduces the backgrounds and allows to strongly constrain the model.
We focused on a benchmark point with the scale of the extra dimensions $m_{KK} = 400$ GeV, which means that the mass of the $(1,1)$ state is close to $600$ GeV.
In this specific point, with an integrated luminosity of 1/fb (10/fb), the significance of the signal exceeds 5 for a branching ratio of the decay of the $(1,1)$ vector photon into top quarks above 13\% (5\%).
This implies that the effective cross-section for discovery is above $210$ fb ($36$ fb).

This analysis can be also extended to 4-tops coming from the tiers $(2,0)$ and $(0,2)$: for $m_{KK} = 300$ GeV (for which the masses are similar to the masses considered in this study), the contribution of a single tier to the 4-top final state is $77$ fb ($144$ fb if both tiers contribute), thus this events can be accounted for by an effective $b = 7.7\%$ ($11\%$).
Using the results of our analysis in this case, we can conclude that this signal can be discovered with an integrated luminosity above $3.5$/fb ($1.5$/fb).

On more general grounds this kind of analysis can be applied to other extra-dimensional models or to models of electroweak symmetry breaking in which new particles strongly interact with the top quarks.


\section*{Acknowledgements}

We thank J.Llodra--Perez for useful discussions and comments.



\appendix

\section{Appendix: Calculation of the mass splitting in tier $(1,1)$}
\label{app:calcul}
\setcounter{equation}{0}
\setcounter{footnote}{0}

The calculation of the mass splitting for tier $(1,1)$ can be performed in a similar way to the other levels.
Details about the calculation can be found in~\cite{Cacciapaglia:2009pa,jeremie}: the main technical difference is that, contrary to tiers $(1,0)$ and $(0,1)$ considered in~\cite{Cacciapaglia:2009pa}, states in this tier carry momentum along both extra directions, thus the calculation is terms of 5D propagators is more involved.

For a vector gauge boson of SU(N), gauge loops contribution to the mass of a generic tier $(n,m)$ with $n+m$ even (thus including the tier $(1,1)$) can be written as:
\begin{multline}
\left. \delta m^2_{V^{(n,m)}} \right|_{\rm gauge} = \frac{\alpha C_2 (G)}{16 \pi^3 R^2} \left[ T_6 + p_g \left( 14 \zeta(3) + (-1)^m B_{Ge} (n,m)  \right) + \right. \\
\left. p_g p_r \left( 14 \zeta(3) + (-1)^n B_{Ge} (m,n)  \right) + p_r 16 \pi^2 (n^2+m^2) \log \Lambda^2 R^2 \right]\,,
 \end{multline}
where $p_r$ and $p_g$ are the parities of the gauge boson in the loop under rotation and glide respectively (in our case, 
they are both $p_{r,g}=+1$), $\alpha = g^2/4\pi$ is the gauge coupling and $C_2 (G) = N$ a gauge group factor.
The other parameters are $T_6 \approx 1.92$, which is a number coming from the renormalised contribution of the torus to the 
loop, the Riemann function $\zeta(3) \approx 1.20$ and the function $B_{Ge}$ which is a complicated function of the KK numbers 
of the tier under consideration.
This function corresponds to a non-local contribution in the bulk, and it gives a very small contribution to the mass and it decreases very fast for large $n$ and $m$: in the case of interest, 
$B_{Ge} (1,1) \approx 1.60$.
The last term of the correction gives the dominant contribution to the mass because it is proportional to the log of the cut-off scale 
$\Lambda$ and to the mass of the tier $n^2+m^2$: in the numerical studies we will also fix $\Lambda R = 10$, however the 
results are mildly dependent on the precise value of the cut-off.
This last term can also be predicted by using the counter-term method described in~\cite{Cacciapaglia:2011hx} without performing 
any loop calculation. The masses will also receive a contribution from the scalar and fermion loops:
\begin{multline}
\left. \delta m^2_{V^{(n,m)}} \right|_{\rm matter} = 
\frac{\alpha}{16 \pi^3 R^2} \left[ \sum_{\rm fermions} C (r_f) \left(  - 8 T_6  \right)  + \sum_{\rm scalars} C (r_s) \left( T_6 + \phantom{\frac{2}{3}} \right.\right.\\
\left. \left.  p_g \left( 7 \zeta(3) + (-1)^m B_{Se} (n,m)  \right) +
p_g p_r \left( 7 \zeta(3) + (-1)^n B_{Se} (m,n)  \right) - p_r \frac{2}{3} \pi^2 (n^2+m^2) \log \Lambda^2 R^2  \right)\right]\,,
 \end{multline}
where $C(r_s)$ and $C (r_f)$ are group factors for scalars $s$ and fermions $f$ (for a fundamental of SU(N), $C (r) = 1/2$, while 
for a U(1) gauge boson $C$ is replaced by the charge squared $Y_{s,f}^2$).
Also, $B_{Se} (1,1) \approx 1.89$. A similar calculation can be performed for fermion fields.
At tree level, there are 2 degenerate fermions: loop corrections distinguish the two because of different couplings, therefore we 
can define as fermion $a$ the fermion that receives the divergent contribution of the loop~\cite{Cacciapaglia:2011hx}, and $b$ the 
one that receives only finite contributions. Note that this is not exactly a mass eigenstate, because there are still small finite terms, 
dependent on the KK numbers, that mix the two states. For simplicity, we will neglect these contributions (which are numerically 
sub-leading) and consider $a$ and $b$ as mass eigenstates. The loop correction to a fermion from gauge loops will have the 
structure:
\begin{equation}
\left. \delta m_{\psi_a^{(n,m)}} \right|_{\rm gauge} = \frac{\alpha C_2 (r_\psi)}{16 \pi^3 R \sqrt{n^2+m^2}} \left[ p_g\; \frac{21}{2} 
\zeta(3) + p_g p_r \; \frac{21}{2} \zeta(3) +  p_r 8 \pi^2 (n^2+m^2) \log \Lambda^2 R^2 \right]\,,
 \end{equation}
where the gauge factor $C_2 (r_\psi)$ is equal to $\frac{N^2-1}{2N}$ for a fundamental of SU(N) and $Y_\psi^2$ for a U(1) field.
The third generation quarks, top singlet and doublet, will also receive sizable corrections from the Yukawa couplings:
\begin{equation}
\left. \delta m_{\psi_a^{(n,m)}} \right|_{\rm Yukawa} = \frac{y_\psi^2}{64 \pi^4 R\sqrt{n^2+m^2}} \left[ p_g\; \frac{21}{4} \zeta(3) + p_g p_r \; \frac{21}{4} \zeta(3) +  p_r \pi^2 (n^2+m^2) \log \Lambda^2 R^2 \right]\,,
 \end{equation}
where $p_r$ and $p_g$ are the parities of the Higgs boson ($p_{r,g} = +1$ in our case of interest).
Note that for the fermion $\psi_b$ the same formula applies except for the divergent term which is absent; moreover, there are some finite contributions that depend on $n$ and $m$ and have non-diagonal pieces.
Here we neglect them altogether due to their small numerical value.
Numerical values of the corrections, in units of $R$, are listed in Table~\ref{tab:loopG} and~\ref{tab:loopF} for all the SM fields.

\begin{table}[tb]\begin{center}
\begin{tabular}{lccc}
\toprule
  & U(1) & SU(2) &  SU(3) \\
\midrule
$\delta m^2_{A_\mu^{(1,1)}} R^2$ & $-0.004$ & $0.168$ & $0.891$\\
$\delta m^2_{A_\phi^{(1,1)}} R^2$ & $-0.0028$ & $0.0064$ & $0.0076$\\
\bottomrule     
\end{tabular}  
\caption{Numerical value of the loop corrections for gauge vector and scalar bosons in tier $(1,1)$, in units of $R$.}
\label{tab:loopG}
\end{center} \end{table}

\begin{table}[tb]\begin{center}
\begin{tabular}{lccccccc}
\toprule
  & $l_R$ & $l_L$,  $\nu_l$ & $u_s$ & $d_s$ & $u_d$, $d_d$ & $t_s$ & $b_d$, $t_d$ \\
\midrule
$\delta m_{f_a^{(1,1)}} R$ & $0.009$ & $0.026$ & $0.150$ & $0.146$ & $0.69$ & $0.159$ & $0.179$ \\
$\delta m_{f_b^{(1,1)}} R$ & $0.0037$ & $0.0010$ & $0.0058$ & $0.0057$ & $0.0066$  & $0.0072$ & $0.0080$ \\
\bottomrule     
\end{tabular}  
\caption{Numerical value of the loop corrections for fermions in tier $(1,1)$, in units of $R$.}
\label{tab:loopF}
\end{center} \end{table}

The tier $(1,1)$ also contains a resonance of the Higgs scalar.
However, the mass correction is quadratically sensitive to the cut-off of the theory: this implies on one hand that the 
calculation is not reliable, and on the other that the corrections tend to be large, therefore the Higgs will be heavier than the 
other states in the tier. The quadratic divergence appears at two levels: there is a bulk contribution and a localised one, characterised 
by two independent counter-terms. While the bulk contribution is the same for the zero mode and the resonances, the localised 
one contributes differently due to the different normalisations of the wave functions of the states.
For instance, the scalar in tier $(1,1)$ has a factor of $2$ in the normalisation with respect to the zero mode, therefore the 
localised divergence will be $4$ times larger.
The divergent loop contribution to the zero mode can be reabsorbed in the definition of the Higgs vacuum expectation value (and 
therefore it contributes to the hierarchy problem of the Higgs sector): if the divergent term were the same for all the resonances, it 
would be possible to rewrite them in terms of the Higgs mass. However, this is not the case, and an additional quadratic 
dependence on the cut-off scale remains in the Higgs resonance masses.

\section{Appendix: Decays of the lightest state in the tier $(1,1)$}
\label{app:locops}
\setcounter{equation}{0}
\setcounter{footnote}{0}

The decays of the lightest state in the tier $(1,1)$ are necessarily mediated by operators localised on the singular points, 
and with different coefficient on the two points thus breaking the additional KK parity of the bulk.
If we consider the lowest order operators, we have kinetic terms that have dimension 6 in unit of mass when localised (this is coming from the canonical dimension of the bulk fields).
However, the gauge scalars do not appear, because they are odd under the rotation symmetry of the orbifold, as 
they have wave functions
\beq
A_5 = \frac{1}{\sqrt{2} \pi R} \sin  \frac{x_5}{R} \cos \frac{x_6}{R} A_\phi^{(1,1)}\,, \qquad A_6 = -\frac{1}{\sqrt{2} \pi R} \cos  \frac{x_5}{R} \sin \frac{x_6}{R} A_\phi^{(1,1)}\,,
\eeq
which vanish on the fixed points.
In fact, on the singular points, both energy-stress tensor $F_{MN}^i = \partial_M A_N^i - \partial_N A_M^i + g f^{ijk} A^j_M A^k_N$ and covariant derivative $D_M = \partial_M - i g A_M^i T^i$ do not contain the gauge scalars:
\beq
F_{5\mu} & = & F_{6\mu} = 0\,,\\
F_{56} &=& \frac{\sqrt{2}}{\pi R^2} \sin \frac{x_5}{R} \sin \frac{x_6}{R} A_\phi^{(1,1)} = 0\,,\\
D_{5,6} \phi &=& \partial_{5,6} \phi\,.
\eeq
In order to have operators which contain $A_\phi^{(1,1)}$ on the fixed points, we need to consider higher derivatives along the two extra coordinates.
For instance, on the fixed points:
\beq
\partial_5 A_5 &=& \frac{1}{\sqrt{2} \pi R^2} \cos \frac{x_5}{R} \cos \frac{x_6}{R} A_\phi^{(1,1)} =  \frac{1}{\sqrt{2} \pi R^2} A_\phi^{(1,1)} \,, \\
\partial_6 A_6 &=& -\frac{1}{\sqrt{2} \pi R^2} \cos \frac{x_5}{R} \cos \frac{x_6}{R} A_\phi^{(1,1)} =  -\frac{1}{\sqrt{2} \pi R^2} A_\phi^{(1,1)} \,.
\eeq
Such terms can be generated by a double covariant derivative applied on an even field: for instance $D_5 D_5 \phi$.
Analogously
\beq
\partial_5 F^i_{5\mu} & = &- \frac{1}{\sqrt{2} \pi R^2} \partial_\mu A_\phi^{i,(1,1)} + \frac{g}{\sqrt{2} \pi R^2} f^{ijk} A_\phi^{j,(1,1)} A_\mu^{k,(0,0)}\,, \\
\partial_6 F^i_{6\mu} & = & \frac{1}{\sqrt{2} \pi R^2} \partial_\mu A_\phi^{i,(1,1)} - \frac{g}{\sqrt{2} \pi R^2} f^{ijk} A_\phi^{j,(1,1)} A_\mu^{k,(0,0)}\,,
\eeq
where we only consider the zero mode for the vector component and $g = g_6/(2 \pi R)$.
Finally, we can consider
\beq
\partial_5 \partial_6 F_{56} = \frac{\sqrt{2}}{\pi R^4} A_\phi^{(1,1)}\,.
\eeq
Note however that the dimensions in mass of such terms are very different:
\beq
D_5 D_5\,, \quad D_6 D_6 &\rightarrow& \mbox{dim.}\; 2\,, \nonumber\\
D_5 F_{5\mu}\,, \quad D_6 F_{6\mu} &\rightarrow& \mbox{dim.}\; 4\,, \nonumber\\
D_5 D_6 F_{56}\,, \quad D_6 D_5 F_{56} &\rightarrow& \mbox{dim.}\; 5\nonumber\,.
\eeq
These can be used to build the lowest order operators that contain $A_\phi^{(1,1)}$.

\subsubsection*{Couplings to zero mode fermions}

The lowest order operators have dimension in mass 8:
\beq
\bar{\psi} \Gamma^\mu (D_\mu D_5 D_5 \psi) + h.c.\,, \quad \bar{\psi} \Gamma_{5,6} (D_5 D_5 D_{5,6} \psi) + h.c.\,,
\eeq
and variations in the position of the derivatives and index $5$--$6$.
Note also that the second operator vanishes on the zero modes because of their chiral nature.
Other possible operators of dimension 9 are:
\beq
H \bar{\psi} D_5 D_5 \psi + h.c.\,, \quad D_5 F_{5\mu} \bar{\psi} \Gamma^\mu \psi\,.
\eeq

\subsubsection*{Couplings to zero mode gauge vectors}

The only dimension 8 operator (for a non Abelian gauge group only) is:
\beq
(D_\mu D_5 F_{5\nu})^i F^{i, \mu \nu} = \left[ (\partial_\mu \partial_5 F^i_{5\nu} + g_6 f^{ijk} A^j_\mu \partial_5 F^k_{5\nu} \right] F^{i,\mu\nu}\,.
\eeq
And similar one with index $6$. However, when extracting the coupling $A_\phi^{(1,1)} A_\mu^{(0,0)} A_\nu^{(0,0)}$, we notice 
that both contributions on the right-handed part of the previous formula vanish: therefore such operators do not contribute to the 
decay of the gauge scalar into a pair of SM gauge bosons.

\subsubsection*{Couplings to zero mode Higgs}

If we consider scalar fields, there is an operator with dimension in mass 6:
\beq
H^\dagger (D_5 D_5 H) + h.c. = - i g_6 (\partial_5 A_5^i) H^\dagger T^i H\,.
\eeq
And similarly with $6$. This operator mediates the decay of $A_\phi^{(1,1)}$ to a pair of Higgs bosons and its decay width 
will be discussed in the following.

\subsection{Decay widths}

Here, we focus on the concrete case of a U(1) gauge scalar coupled to fermions (tops) and scalars.
We will also consider the massive vector as a comparison.
The relevant operators are:
\beq
\mathcal{L}_{\rm loc} (A_\mu) & = & i \frac{z}{\Lambda^2} \bar{\psi} \Gamma^\mu (\partial_\mu - i g_6 Q_f A_\mu) \psi\,, \\
\mathcal{L}_{\rm loc} (A_\phi) &=& i \frac{\zeta}{\Lambda^2} H^\dagger D_5 D_5 H 
+ i \frac{\xi}{\Lambda^4} \bar{\psi} \Gamma^\mu (D_\mu D_5 D_5 \psi) + h.c.
\eeq
where we do not consider other combinations of derivatives.
Extracting the coupling of one $A_\phi^{(1,1)}$ with zero modes, we obtain:
\beq
\mathcal{L}_{\rm loc} (A^{(1,1)}_\mu) & = & \frac{z_{L,R}}{2 \pi^2 R^2 \Lambda^2} g Q_f \bar{\psi}_{L,R}^{(0,0)} \gamma^\mu 
P_{L,R} \psi_{L,R}^{(0,0)} A_\mu^{(1,1)} + \dots\,, \\
\mathcal{L}_{\rm loc} (A^{(1,1)}_\phi) &=& \frac{\zeta}{2 \pi^2 R^2 \Lambda^2}  \frac{g Q_h}{\sqrt{2}} H^{(0,0) \dagger} H^{(0,0)} 
A^{(1,1)}_\phi\nonumber\\
&+&  \frac{\xi_{L,R}}{2 \pi^2 R^4 \Lambda^4} \frac{g Q_f R}{\sqrt{2}} \bar{\psi}_{L,R}^{(0,0)} \gamma^\mu 
\psi_{L,R}^{(0,0)} \partial_\mu A_\phi^{(1,1)}+ h.c. + \dots
\eeq
The $L,R$ refer to the chirality of the zero mode fermion. From these we can calculate the partial widths
\beq
\Gamma (A_\mu^{(1,1)} \to \bar{f} f) &=& \frac{\alpha Q_f^2 M}{6} \frac{1}{4 \pi^4 (\Lambda R)^4} \sqrt{1-\frac{4m_f^2}{M^2}} \left( (z_L^2 + z_R^2) \left( 1- \frac{m_f^2}{M^2} \right) + 6 z_L z_R \frac{m_f^2}{M^2} \right)\,. \\
\Gamma (A_\phi^{(1,1)} \to \bar{f} f ) &=&  \frac{\alpha Q_f^2 M}{8} \frac{1}{4 \pi^4 (\Lambda R)^8} (m_f R)^2  \sqrt{1-\frac{4m_f^2}{M^2}}  (\xi_L - \xi_R)^2 \,. \\
\Gamma (A_\phi^{(1,1)} \to h h ) &=& \frac{\alpha Q_h^2}{16 R^2 M} \frac{\zeta^2}{4 \pi^4 (\Lambda R)^4} \sqrt{1-\frac{4m_f^2}{M^2}}\,.
\eeq

To have an idea of the rates, we calculated the partial decay width for $M = \sqrt{2}\cdot\, 400\; \mbox{GeV} =
565$ GeV and $z_L = z_R = \xi_L = -\xi_R = \zeta = 1$. Moreover, we used $m_t = 175$ GeV, $m_h = 120$ GeV, $\Lambda R = 10$, $\alpha = \alpha_{em} = 1/127$, $Q_f=Q_h=1$:
\beq
\Gamma (A_\mu^{(1,1)} \to \bar{f} f) &=& 3.6 \times 10^{-7}\; \mbox{GeV}\,. \\
\Gamma (A_\phi^{(1,1)} \to \bar{f} f ) &=&  1.7 \times 10^{-11}\; \mbox{GeV} \,. \\
\Gamma (A_\phi^{(1,1)} \to h h ) &=& 3.2 \times 10^{-8}\; \mbox{GeV}\,.
\eeq
It is therefore natural to expect that the branching of the gauge scalar in a pair of tops is of order $10^{-3}\div 10^{-4}$.
This implies that the gauge scalars do not contribute significantly to the 4-top final state; on the other hand, the fact that the only dimension-6 operator involves Higgs bosons implies that the branching ratio into $h h$ is naturally 100\%.
This would lead to very large production cross sections of 4 Higgs bosons and interesting final states that may lead to a discovery of the Higgs and of the extra-dimensional model.
For instance, for a light Higgs, one may select  a final state with 6 bottoms and a pair of photons, with a total branching of $4 \cdot BR(h\to\gamma \gamma) \sim 10^{-2}$ and an effective cross-section of the order of $100$ fb.




\begin{thebibliography}{99}

\bibitem{Cheung:1995eq}
  K.~m.~Cheung,
  arXiv:hep-ph/9507411.
  
\bibitem{Spira:1997ce}
  M.~Spira and J.~D.~Wells,
  Nucl.\ Phys.\  B {\bf 523} (1998) 3
  [arXiv:hep-ph/9711410].
  
\bibitem{Hill:2002ap}
  see e.g., C.~T.~Hill, E.~H.~Simmons,
  Phys.\ Rept.\  {\bf 381 } (2003)  235-402.
  [hep-ph/0203079], and references therein.
  
\bibitem{Pomarol:2008bh}
  A.~Pomarol and J.~Serra,
  Phys.\ Rev.\  D {\bf 78} (2008) 074026
  [arXiv:0806.3247 [hep-ph]].

\bibitem{Kumar:2009vs}
  K.~Kumar, T.~M.~P.~Tait and R.~Vega-Morales,
  JHEP {\bf 0905} (2009) 022
  [arXiv:0901.3808 [hep-ph]].

\bibitem{Frigerio:2011zg}
  M.~Frigerio, J.~Serra and A.~Varagnolo,
  JHEP {\bf 1106} (2011) 029
  [arXiv:1103.2997 [hep-ph]].
     
\bibitem{Kane:2011zd}
  G.~L.~Kane, E.~Kuflik, R.~Lu and L.~T.~Wang,
  arXiv:1101.1963 [hep-ph].
  
\bibitem{Jung:2010ms}
  S.~Jung and J.~D.~Wells,
  JHEP {\bf 1011} (2010) 001
  [arXiv:1008.0870 [hep-ph]].
  
  \bibitem{Cacciapaglia:2009pa}
  G.~Cacciapaglia, A.~Deandrea and J.~Llodra-Perez,
  JHEP {\bf 1003} (2010) 083
  [arXiv:0907.4993 [hep-ph]].

\bibitem{LSP}
  G.~Jungman, M.~Kamionkowski and K.~Griest,
  Phys.\ Rept.\  {\bf 267} (1996) 195
  [arXiv:hep-ph/9506380];
G.~Bertone, D.~Hooper and J.~Silk,
  Phys.\ Rept.\  {\bf 405} (2005) 279
  [arXiv:hep-ph/0404175].

\bibitem{LTP}
  I.~Low,
  JHEP {\bf 0410} (2004) 067
  [arXiv:hep-ph/0409025].
  
\bibitem{UED5}
  T.~Appelquist, H.~C.~Cheng and B.~A.~Dobrescu,
  Phys.\ Rev.\  D {\bf 64} (2001) 035002
  [arXiv:hep-ph/0012100].
  
\bibitem{LKP}
  G.~Servant and T.~M.~P.~Tait,
  Nucl.\ Phys.\  B {\bf 650}, 391 (2003)
  [arXiv:hep-ph/0206071].

\bibitem{Hebecker:2003we}
  A.~Hebecker,
  JHEP {\bf 0401} (2004) 047
  [arXiv:hep-ph/0309313].

\bibitem{raviolo}
  B.~A.~Dobrescu and E.~Ponton,
  JHEP {\bf 0403} (2004) 071
  [arXiv:hep-th/0401032];

  E.~Ponton and L.~Wang,
  JHEP {\bf 0611} (2006) 018
  [arXiv:hep-ph/0512304].



\bibitem{Dohi:2010vc}
  H.~Dohi and K.~y.~Oda,
  Phys.\ Lett.\  B {\bf 692} (2010) 114
  [arXiv:1004.3722 [hep-ph]].
  
  
 \bibitem{jeremie}
J.~Llodra-Perez, 
``Effective Models of New Physics at the Large Hadron Collider'',
PhD Thesis of the University Claude Bernard (Lyon 1), 2011. 
  
  
 \bibitem{relicDM} 
  G.~Cacciapaglia, A.~Deandrea and B.~Kubik-Deriaz,
  work in preparation.   
 
\bibitem{Burdman:2006gy}
  G.~Burdman, B.~A.~Dobrescu and E.~Ponton,
  Phys.\ Rev.\  D {\bf 74} (2006) 075008
  [arXiv:hep-ph/0601186].

 
  
\bibitem{Cacciapaglia:2011hx}
  G.~Cacciapaglia, A.~Deandrea and J.~Llodra-Perez,
  arXiv:1104.3800 [hep-ph].



\bibitem{Alwall:2007st}
  J.~Alwall, P.~Demin, S.~de Visscher, R.~Frederix, M.~Herquet, F.~Maltoni, T.~Plehn, D.~L.~Rainwater {\it et al.},
  JHEP {\bf 0709 } (2007)  028.
  [arXiv:0706.2334 [hep-ph]].
  
  
\bibitem{Christensen:2008py}
  N.~D.~Christensen, C.~Duhr,
  Comput.\ Phys.\ Commun.\  {\bf 180 } (2009)  1614-1641.
  [arXiv:0806.4194 [hep-ph]].

\bibitem{Sjostrand:2006za}
  T.~Sjostrand, S.~Mrenna, P.~Z.~Skands,
  JHEP {\bf 0605 } (2006)  026.
  [hep-ph/0603175].

\bibitem{Pumplin:2002vw}
  J.~Pumplin, D.~R.~Stump, J.~Huston, H.~L.~Lai, P.~M.~Nadolsky and W.~K.~Tung,
  JHEP {\bf 0207} (2002) 012
  [arXiv:hep-ph/0201195].


 \bibitem{Kidonakis:2009mx}
  N.~Kidonakis,
  [arXiv:0909.0037 [hep-ph]].
  


\bibitem{Delphes}
S. Ovyn, X. Rouby, and V. Lemaitre, 
 [hep-ph/0903.2225]
 
 \bibitem{fastjet}
 M.~Cacciari and G. P. Salam, 
 Phys. Lett. B641 (2006) 57-61, 
 [hep-ph/0512210].


 
 
 
\end{thebibliography}
\end{document}